\documentclass[11pt,a4paper]{article}

\usepackage{url}
\usepackage{jcappub}
\usepackage{bm}
\usepackage{amsfonts}
\usepackage{subfigure}
\usepackage{graphicx}
\usepackage{upgreek}

\renewcommand\[{\left[}

\newcommand{\be}{\begin{equation}}
\newcommand{\ee}{\end{equation}}
\newcommand{\bea}{\begin{eqnarray}}
\newcommand{\eea}{\end{eqnarray}}

\newcommand{\beq}{\begin{equation}}
\newcommand{\eeq}{\end{equation}}

\graphicspath{{figure/}}

\begin{document}


\title{Axion-like particles from primordial black holes shining through the Universe} 

\author[a]{Francesco Schiavone,}
\author[b,c]{Daniele Montanino,}
\author[a,d]{Alessandro Mirizzi,}
\author[e]{Francesco Capozzi}

\emailAdd{f.schiavone15@studenti.uniba.it}\emailAdd{alessandro.mirizzi@ba.infn.it}\emailAdd{daniele.montanino@le.infn.it}
\emailAdd{capozzi@vt.edu} 

\affiliation[a]{Dipartimento Interateneo di Fisica ``Michelangelo Merlin'', Via Amendola 173, 70126 Bari, Italy}
\affiliation[b]{Dipartimento di Matematica e Fisica ``Ennio De Giorgi'', Universit\`a del Salento, Via Arnesano, 73100 Lecce, Italy}
\affiliation[c]{Istituto Nazionale di Fisica Nucleare - Sezione di Lecce,
Via Arnesano, 73100 Lecce, Italy.}  
\affiliation[d]{Istituto Nazionale di Fisica Nucleare - Sezione di Bari, Via Orabona 4, 70126 Bari, Italy}
\affiliation[e]{Center for Neutrino Physics, Department of Physics, Virginia Tech, Blacksburg, VA 24061, USA}

\abstract{We consider a cosmological scenario in which  the very early Universe experienced
a transient epoch of matter domination due to the formation of a large population of primordial
black holes (PBHs) with masses  $M \lesssim 10^{9}~\textrm{g}$,
that evaporate
before Big Bang nucleosynthesis. In this context, Hawking radiation would be  a non-thermal mechanism to produce 
a cosmic background of axion-like particles (ALPs). We assume the minimal scenario in which these ALPs couple only with photons.
 In the case of ultralight ALPs ($m_a \lesssim 10^{-9}$~eV)  the cosmic magnetic fields might trigger
ALP-photon conversions, while for masses $m_a \gtrsim 10$~eV spontaneous ALP decay in photon pairs would be effective.
We investigate the impact of these mechanisms on the cosmic X-ray background, on the excess in X-ray luminosity in Galaxy Clusters, and on the process of cosmic reionization.}
 
\maketitle

\section{Introduction}

It has been proposed a long time ago that primordial black holes (PBHs) could have been produced by random density
fluctuations in the very early Universe~\cite{zeld,Hawking:1971ei,Carr:1974nx,Carr:2005zd}.
These objects would form when a mass of cosmic fluid came to be contained within its Schwarzschild
radius. One finds that the mass
of a PBH can be related to its time of formation $t_f$ by~\cite{zeld}
\begin{equation}
\label{eq:bh_initial_mass}
	M \sim m_P^2 t_f \sim 10^6\left(\frac{t_f}{10^{-32} \,\textrm{s}} \right) \textrm{g} \,\ ,
\end{equation}
where $m_P=1/\sqrt{G}=1.22 \times 10^{19}$~GeV is the Planck mass.
The density perturbations from which PBHs formed would have arisen only after the 
end of inflation~\cite{GarciaBellido:1996qt,Kawasaki:2016pql,Kannike:2017bxn,Kawasaki:1997ju,Hsu:1990fg,La:1989za,Polnarev:1986bi}. 
On the other hand, 
for masses $M \lesssim 10^{9} \,\textrm{g}$ 
PBHs would \emph{evaporate} via \emph{Hawking radiation} 
 before the start of Big Bang Nucleosynthesis (BBN)~\cite{Papanikolaou:2020qtd},
 being essentially unconstrained by current observations.
 Nevertheless, in the mass range ${10}~\textrm{g}\lesssim M \lesssim 10^{9}~\textrm{g}$,
 PBHs might cause a matter dominated epoch in the early Universe.

A black hole emits particles via Hawking radiation almost as if it were a blackbody at a temperature defined by~\cite{Hawking:1974sw}
\begin{equation}
\label{eq:bh_temp}
	T_{\rm BH}= \frac{m_P^2}{8 \pi M} \simeq 10^7 \left(\frac{10^6 \,\textrm{g}}{M} \right) \,\textrm{GeV} \,\ .
\end{equation}
One of the most significant facts about Hawking radiation is that  PBHs emit 
\emph{``democratically''}  {any} particle (regardless of its spin or coupling) whose rest mass is below $T_{\rm BH}$.
Bounds on photon, neutrino, positron emission have been widely discussed (see, e.g.~\cite{Carr:2020gox,Arbey:2019vqx,Dasgupta:2019cae, Wang:2020uvi,
Calabrese:2021zfq, Lunardini:2019zob,DeRocco:2019fjq,Iguaz:2021irx,Mittal:2021egv}).
Furthermore, novel particles in the dark sector with feeble couplings with Standard Model sector might also be emitted.
In this context,
the case of gravitons has been widely explored (see, e.g.,~\cite{Dolgov:2011cq,Inomata:2020lmk,Hooper:2020evu}). 
In general, the evaporation of primordial black holes has been considered as  a particularly attractive mechanism for the production of both dark radiation and dark matter 
(see, e.g.,~\cite{Morrison:2018xla,Fujita:2014hha,Lennon:2017tqq,Hooper:2019gtx,Baldes:2020nuv,Hooper:2020evu,Masina:2020xhk,Auffinger:2020afu,Gondolo:2020uqv,Masina:2021zpu,Cheek:2021odj,Arbey:2021ysg}). In this regard, in~\cite{Hooper:2019gtx} it has been shown that Hawking radiation of  particles with masses 
$m_a \lesssim 10$~MeV would behave as dark radiation, {significantly extending the mass range expected for thermally produced dark radiation.}

In this context, we devote our work to study the physics case of axion-like 
particles (ALPs)~\cite{Ringwald:2014vqa} emitted from PBHs. 
Hawking evaporation would represent a novel non-thermal mechanism to produce a cosmic background of ALPs
(see~\cite{Dror:2021nyr} for a recent investigation of different cosmic ALP backgrounds).
In the recent literature, other non-thermal mechanisms have been proposed to produce ALP dark radiation. Notably, 
primordial decays of string theory moduli constitute an intriguing scenario~\cite{Conlon:2013isa} 
leading to interesting phenomenological
consequences that have been widely explored in literature~\cite{Conlon:2013txa,Angus:2013sua,Conlon:2015uwa,Day:2015xea,Fairbairn:2013gsa,Evoli:2016zhj}. 
We extend these results to the case of ALP Hawking radiation, focussing on PBH masses $M \lesssim 10^{9}~\textrm{g}$.
We assume a minimal scenario, in which ALPs only have a two-photon coupling $g_{a\gamma}$. 
Depending on the mass of the ALPs, different phenomena would arise.
Notably, for ultralight ALPs ($m_a \lesssim 10^{-9}$~eV)  the cosmic magnetic fields might trigger
ALP-photon conversions leading to a diffuse X-ray flux. On the other hand, for higher masses
 the conversions in the magnetic fields would be suppressed, while for $m_a \gtrsim 10$~eV
 spontaneous ALP decay in photon pairs would be effective. 
 We use different observables discussed in previous literature to constrain the ALP-induced X-ray flux. 
 In particular, we consider the measurement of the cosmic X-ray background, the
 excess in X-ray luminosity in Galaxy Clusters, and   the effect of ALPs on cosmic reionization.

The plan of our work is as follows. In Sec.~2 we characterize the ALP emission from PBHs via Hawking radiation obtaining
 the ALP number density spectrum. In Sec.~3 we discuss the ALP-photon conversions in magnetic fields, with particular emphasis
 on the case of turbulent primordial magnetic field. We present a perturbative analytical expression useful to characterize
 the photon-ALP conversions in the presence of photon absorption in the early Universe. 
 In Sec.~4 we discuss signatures and constraints on ultralight ALPs from the produced X-ray fluxes after conversions in 
 cosmic magnetic fields. 
 We devote Sec.~5 to specifically analyze the effects of ALP-photon conversions in the early Universe on reionization.
 In Sec.~6 we extend our previous results to the case of massive ALPs decaying into photons. 
 Finally, in Sec.~7 we summarize our results and we conclude.


\section{ALP emission from primordial black holes}

\subsection{Hawking radiation }

\begin{figure}[!t]
 	\centering
 	\includegraphics[width=.8\textwidth]{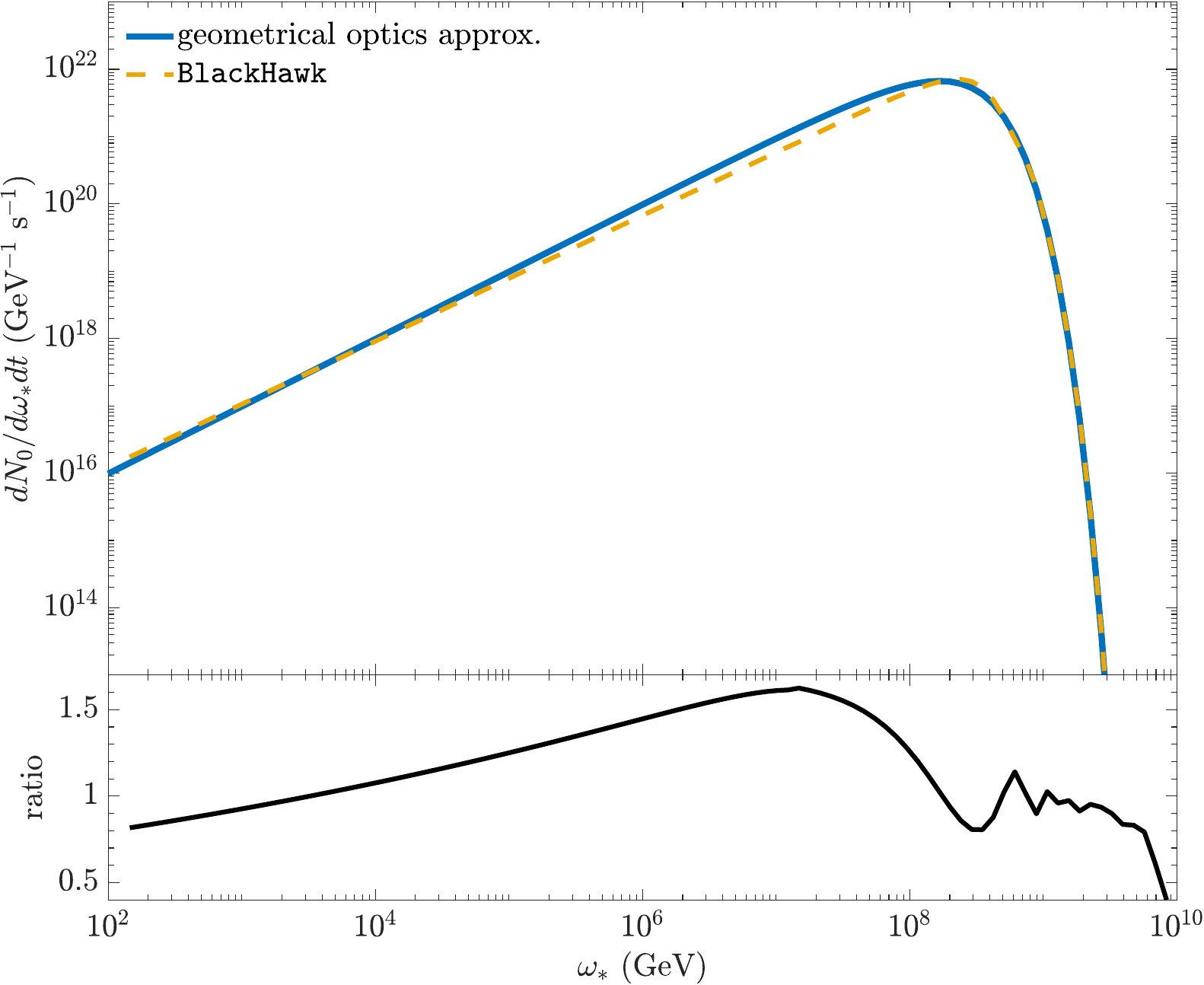}
 	\caption{Instantaneous spectrum of a spin-0 particle emitted by a black hole with sample mass $M=10^5\,{\rm g}$ in the geometrical optics approximation, compared to the numerical spectrum computed by {\tt BlackHawk}. The ratio between the spectrum obtained in the geometrical optics approximation
	and that from {\tt BlackHawk} is shown in the bottom panel.}
 	\label{fig:geometrical_vs_blackhawk}
 \end{figure}

The emission rate of a particle state of spin $s$ 
and energy $\omega$ from  a nonrotating, uncharged PBH of mass $M$
is given by~\cite{Hawking:1974rv}
\begin{equation}
\label{eq:hawking_rate}
	\frac{d{N_s}}{d\omega dt}\equiv\phi_s(\omega,M)=\frac{1}{2\pi}\frac{\Gamma_s(\omega,M)}{e^{\omega/T_{\rm BH}}-(-1)^{2s}}\,,\ 
\end{equation}
where $T_{\rm BH}$ is the black-hole temperature defined in Eq.~\eqref{eq:bh_temp} and 
 the function $\Gamma_s(\omega,M)$ is the so-called \emph{greybody factor} that represents a distortion of the ordinary blackbody spectrum corresponding to the probability that emitted particles are reabsorbed into the hole~\cite{Page:1976df,Page:1976ki,Page:1977um,MacGibbon:1990zk}. 
We  stress that the greybody factors have in general a complex dependence from energy. In particular, in the high-energy limit ($\omega\gg T_{\rm BH}$), the greybody factors have an oscillatory behavior, although on average they tend to the optical-geometric approximation $\overline\Gamma_s(\omega,M)\sim 27M^2\omega^2/m_P^4$ for each degree of freedom~\cite{Page:1976df}. Since we will focus on ALPs, for simplicity we can write 
\begin{equation}
\label{eq:gamma_approx}
	\Gamma_s(\omega,M)\simeq \frac{\kappa_a M^2\omega^2}{m_P^4}=\frac{\kappa_a}{(8\pi)^2}\left(\frac{\omega}{T_{\rm BH}}\right)^2\,.\ 
\end{equation}
We computed the factor $\kappa_a$  as the ratio between the total flux of scalars (obtained by using the {\tt BlackHawk} public code \cite{Arbey:2019mbc}) and the flux obtained integrating Eq.~\eqref{eq:hawking_rate} with the approximation in Eq.~\eqref{eq:gamma_approx}, thus finding
\begin{equation}
\kappa_a=\frac{64\pi^3}{T_{\rm BH}\zeta(3)}\int d\omega \, \left(\frac{d{N_0}}{d\omega d{t}}\right)_{\rm BlackHawk}
\simeq 27.6
\, .
\end{equation}
In this way the total flux of scalars calculated with Eqs.~\eqref{eq:hawking_rate} and \eqref{eq:gamma_approx} coincides with that computed by {\tt BlackHawk}. 
{We note that this result is very close to the simple optical-geometric approximation, since the black-hole temperature is generally very high compared to the mass of the radiated particles.} 
 We have verified that the numerical value calculated in Eq.~\eqref{eq:gamma_approx} is almost constant over all the mass range $m_P\lesssim M\lesssim 10^9$~g. Figure~\ref{fig:geometrical_vs_blackhawk} shows a comparison between instantaneous Hawking spectra of a spin-0 particle, computed using the geometrical optics approximation of Eqs. \eqref{eq:hawking_rate} and \eqref{eq:gamma_approx} and {\tt BlackHawk}, respectively (upper panel).
The ratio between the spectrum obtained using the geometrical optics approximations and the {\tt BlackHawk} result is shown in the lower panel, ranging between $\sim$ 0.5 and 2.
For this Figure a sample black hole mass of $M=10^5\,{\rm g}$ has been chosen, but one can see that numerical agreement between the two methods of calculation is quite good over the whole mass range specified above.~\footnote{We note that the only scalar particle in the present release of {\tt BlackHawk} is the Higgs boson. Therefore the BlackHawk spin-0 spectra are cut off at energies smaller than $m_h=125\,{\rm GeV}$.}

From Eq.~\eqref{eq:hawking_rate} we can obtain the PBH mass loss rate \cite{Page:1976ki}
\begin{equation}
\label{eq:hawking_power-loss}
	\frac{d M}{d t}=-\sum_{s}\int   d\omega \, \omega  \frac{d{N_s}}{d\omega dt} \equiv -\frac{f(M)}{M^2}\, ,
\end{equation}
where $f(M)$ is a function that depends weakly on the BH mass $M$. We note that $f(M)$ is obtained by summing over all particle states radiated by the black hole. Since in the mass range $m_P\lesssim M\lesssim 10^9$~g the black-hole temperature is very high ($T_{\rm BH} \gg{100}$~{GeV}), Hawking radiation contains all possible particle states in the Standard Model (plus one pseudoscalar ALP in our hypothesis). Correspondingly, the value of $f(M)$ is essentially constant in this range, as one may check explicitly by using {\tt BlackHawk}. We thus set $f(M)=m_P^4 f_{\rm ev}$ with $f_{\rm ev}=4.26\times 10^{-3}$ \cite{Auffinger:2020afu}. The latter value accounts for Standard Model particles only, as the contribution of a single extra scalar, of order $\mathcal{O}(10^{-5})$ \cite{Taylor:1998dk}, would be negligibly small in comparison.
We neglect further possible non-standard extra (light or heavy) degrees of freedom such as sterile neutrinos, or supersymmetric particles.

Assuming initial conditions $M(0)=M$~\footnote{As pointed out in~\cite{Hooper:2019gtx}, $M$ can be taken to be the initial mass \emph{after} the processes of accretion and merging have ceased to be efficient.},
from Eq.~\eqref{eq:hawking_power-loss} one gets~\cite{Hooper:2019gtx}
\begin{equation}
\label{eq:bh_mass}
	M(t)=M\left(1-\frac{3 f_{\rm ev} m_P^4 t}{M^3}\right)^{1/3}\equiv M\left(1-\frac{t}{\tau_{\rm BH}}\right)^{1/3}\,,
\end{equation}
where one defines the black-hole lifetime $\tau_{\rm BH}$ as
\begin{equation}
\label{eq:bh_time}
	\tau_{\rm BH} = \frac{1}{3f_{\rm ev}}\frac{M^3}{m_P^4} \simeq 4.16\times10^{-1}\left(\frac{M}{10^9\,\textrm{g}}\right)^3\,\textrm{s}\,.
\end{equation}
Thus, as a PBH emits particles, its mass is reduced until it completely disappears---more appropriately, it \textit{explodes}, since the emission rate becomes increasingly faster as the mass is reduced.

\begin{figure}
 	\centering
 	\includegraphics[width=.8\textwidth]{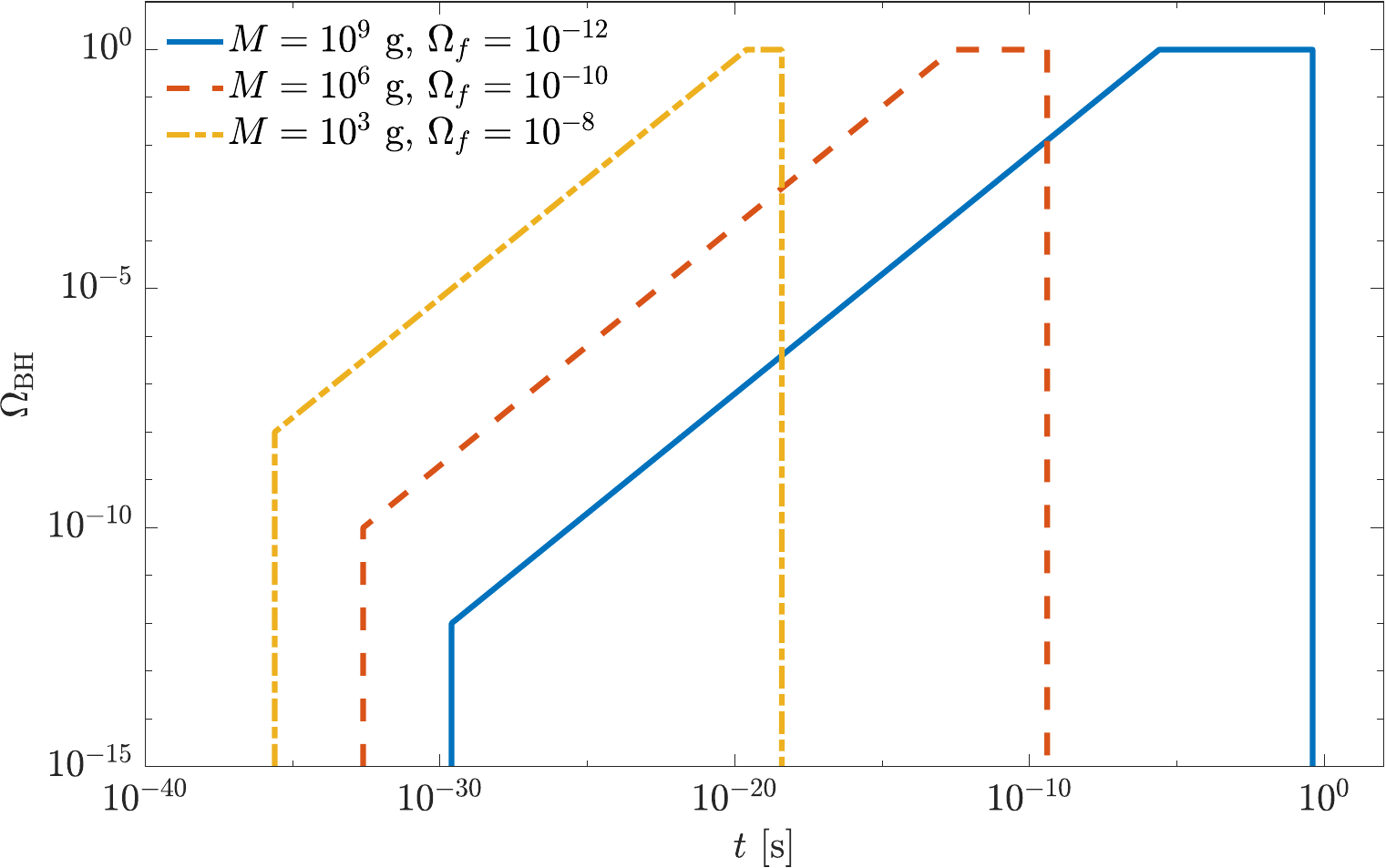}
 	\caption{Evolution of the PBH density parameter as a function of time, for several values of PBH mass and formation density.}
 	\label{fig:pbh_evolution}
 \end{figure}

Eq.~\eqref{eq:bh_time} can be used to place a lower bound on the initial abundance of evaporating PBHs needed to have a PBH-dominated epoch between inflation and BBN. 
Following~\cite{Dolgov:2011cq}, we assume that a population of PBHs with a monochromatic mass spectrum forms in a radiation-dominated epoch with a density parameter $\Omega_f \equiv \Omega_{\rm BH}(t_f) = \rho_{\rm BH}(t_f)/\rho_c\ll 1$. As the Universe expands, it is easy to show that the PBH density grows as $\Omega_{\rm BH}\sim a \sim t^{1/2}$, where $a$ is the scale factor of the Universe. Therefore, the domination time $t_d$ such that $\Omega_{\rm BH}(t_d)=1$ is determined by imposing the condition
 \begin{equation}
 	\Omega_{\rm BH}(t_d)=\Omega_f \left(\frac{t_d}{t_f}\right)^{1/2}=1 \,\ ,
 \end{equation}
 which, using Eq.~\eqref{eq:bh_initial_mass}, results in
 \begin{equation}
 \label{eq:domination_time}
 	t_d=\frac{M}{m_P^2\Omega_f^2}	\simeq 2.49\times 10^{-10}\left(\frac{M}{10^9\,\textrm{g}}\right)\left(\frac{10^{-10}}{\Omega_f}\right)^2\,\textrm{s} \gg t_f\,.
 \end{equation}
 Of course, in order for PBHs to dominate the energy density, their lifetime $\tau_{\rm BH}$ should be longer than $t_d$, i.e. $\tau_{\rm BH} > t_d-t_f\simeq t_d$.
 Using Eqs.~\eqref{eq:bh_time} and \eqref{eq:domination_time}, this condition puts an upper bound on the PBH density parameter in order to have a BH domination
 \begin{equation}
 	\Omega_f\gg(3f_{\rm ev})^{1/2}\,\frac{m_P}{M}=2.43\times10^{-15}\left(\frac{10^9\,\textrm{g}}{M}\right)\,,
 \end{equation}
 that, combined with the BBN upper bound on the PBH mass $M\lesssim 10^9$~g gives
 $\Omega_f\gtrsim 10^{-14}$~\cite{Papanikolaou:2020qtd}. This result confirms that, even if they formed with a very small initial abundance, it is well possible that PBHs  would dominate the energy density of the Universe for a transient period before BBN. {Figure \ref{fig:pbh_evolution} shows the evolution of the PBH density parameter as a function of time, for several values of PBH mass and formation density. It is evident how heavier PBHs form at a later time $t_f$ and have a longer lifetime $\tau_{\rm BH}$, and how PBHs with a lower formation density take a longer time $t_d$ to become dominant.}
 
\subsection{Spectral number density}

For the time being we will focus only on the ALP spectrum. We assume that the ALP mass is always much lower than the black hole temperature during evaporation, $m_a\ll T_{\rm BH}$. From Eq.~\eqref{eq:bh_temp} this condition becomes $m_a\ll 10^4$~GeV in our BH mass range. With this assumption the ALPs are ultrarelativistic during evaporation. The number density of ALPs follows the Boltzmann equation
\begin{equation}
\left[\frac{\partial}{\partial t} -H(t)k\frac{\partial}{\partial k}+2H(t)\right] \frac{dn_a}{dk}(k,t)=n_{\rm BH}(t)\phi_0[k,M(t)]
\, ,
\label{eq:Boltz1}\end{equation}
where $k= |{\bf k}|=\omega$ is the modulus of the ALP momentum, $n_{\rm BH}(t)$ is the number density of PBHs at time $t$ given by $n_{\rm BH}(t)=n_{\rm BH}(t_f)/a^3(t)$ (assuming $a(t_f)=1$), with $n_{\rm BH}(t_f)$ number density of PBHs at the formation time
 \begin{equation}
 	n_{\rm BH}(t_f) =\frac{ \rho_r}{M}\Omega_f=\frac{3 m_P^2}{32\pi Mt_f^2}\Omega_f = \frac{3 m_P^6}{32\pi M^3}\Omega_f\, ,\label{eq:BHdens0}
 \end{equation}
where we have used the fact that in a radiation dominated epoch the critical density $\rho_c = \rho_r$,  the Hubble parameter  $H={\dot a}/a=1/2t$ and the Friedmann equation gives $\rho= 3 m_{\rm P}H^2/8 \pi $. Eq.~\eqref{eq:Boltz1} must be evolved up to evaporation time $t_{\ast}=\tau_{\rm BH}$. The solution can be 
written as in Ref.~\cite{Auffinger:2020afu}
\begin{equation}
\frac{dn_a}{dk}(k_{\ast},\tau_{\rm BH})=n_{BH}^{\ast}\int_{t_f}^{\tau_{\rm BH}}\phi_0\left[k_{\ast}\frac{a_{\ast}}{a(t)},M(t)\right]\frac{a_{\ast}}{a(t)}dt\, ,
\label{eq:spect0}
\end{equation}
where all starred quantities refer to the evaporation time, $a_{\ast}=a(t_{\ast})$, $n_{BH}^{\ast}=n_{\rm BH}(t_f)/a_{\ast}^3$. This equation must be supplemented by the Friedmann-Robertson-Walker and radiation Boltzmann equations
\begin{eqnarray}
H^2(t)=\left(\frac{\dot a}{a}\right)^2&=&\frac{8\pi}{3m_{\rm P}^2}\left[\rho_r(t)+M(t) n_{\rm BH}(t)\right] \,\ ,\nonumber\\
\dot\rho_r(t)+4H(t)\rho_r(t)&\equiv&\frac{1}{a^4(t)}\frac{d}{dt}[a^4(t)\rho_r(t)]=-n_{\rm BH}(t)\dot M(t)
\label{eq:FRW}\, .
\end{eqnarray}

\begin{figure}
\centering
	\includegraphics[width=.8\textwidth]{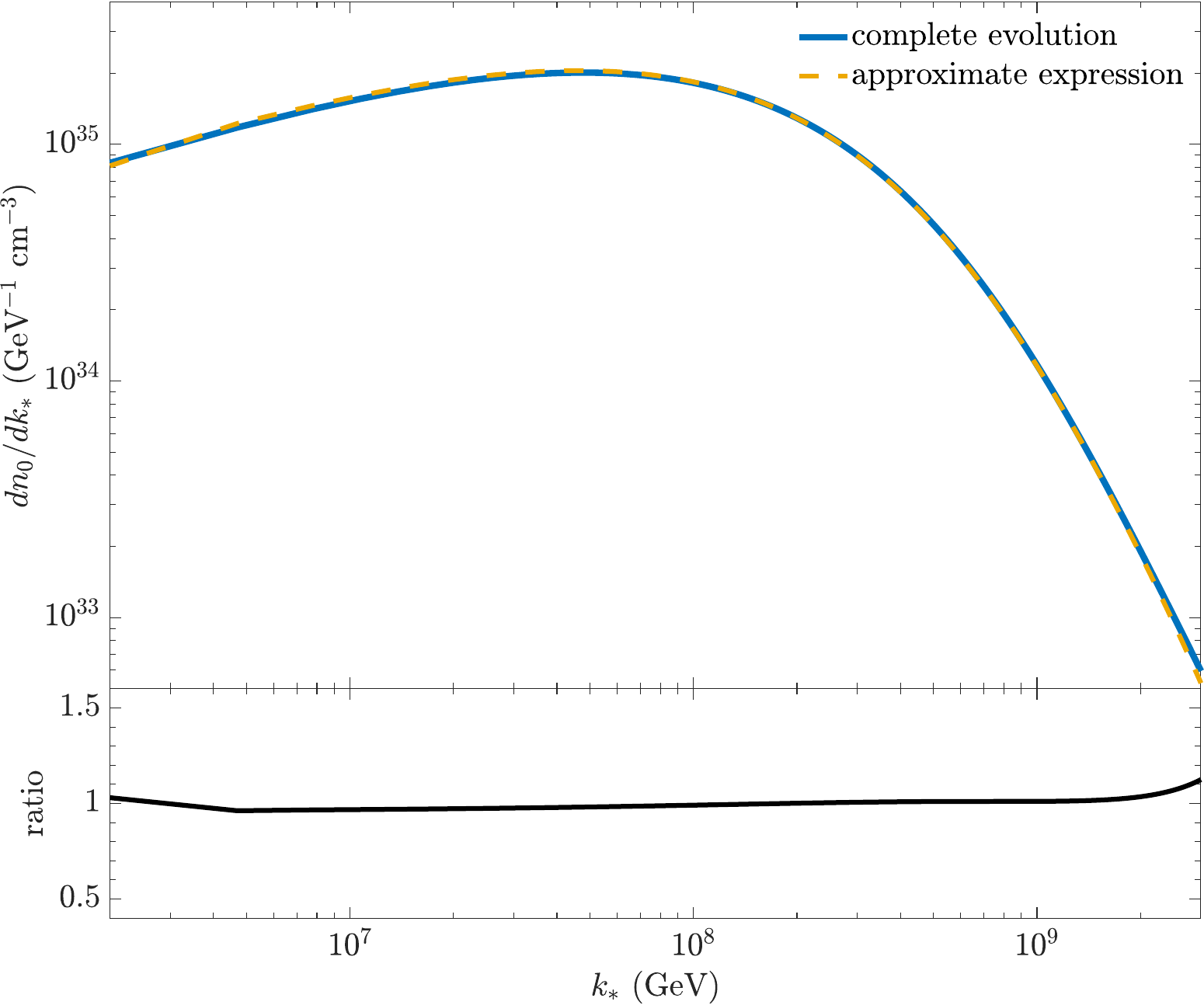}
	\caption{Comparison (upper panel) and ratio (bottom panel) between the present-day scalar Hawking spectra from a population of PBHs of mass $M=10^5\,{\rm g}$ calculated both using the exact cosmological evolution as well as an instantaneous transition to matter domination.}
	\label{fig:comp_vs_approx}
\end{figure}

Previous equations can be simplified assuming that the evolution is essentially BH dominated, that is $t_f<t_d\ll\tau_{\rm BH}$. In this case, in the first of Eqs.~\eqref{eq:FRW} we can neglect $\rho_r$. Moreover, for simplicity we can consider the mass almost constant during the evolution. With this approximation the solution of  Eq.~\eqref{eq:FRW}  is simply
\begin{equation}
a^{3/2}(t)-a^{3/2}(t_d)\simeq\frac{3}{2}\left(\frac{8\pi}{3m_{\rm P}^2}M n_{\rm BH}(t_f)\right)^{1/2} (t-t_d)\, .
\end{equation}
For $t\gg t_d$ we can write
\begin{equation}
\frac{a_{\ast}}{a(t)}\simeq\left(\frac{\tau_{\rm BH}}{t}\right)^{2/3}
\label{eq:at}
\end{equation}
with
\begin{equation}
a_{\ast}\simeq(4f_{\rm ev})^{-2/3}
	\left(\frac{M}{m_P}\right)^{4/3}\Omega_f^{1/3}\,\ .
\label{eq:astar}
\end{equation}
Using this expression we can calculate $n_{\rm BH}^{\ast}$ as 
\begin{equation}
n_{\rm BH}^{\ast}=\frac{n_{\rm BH}(t_f)}{a_{\ast}^3}=
 \frac{3}{2\pi}f_{\rm ev}^2\frac{m_P^{10}}{M^7} \,\ .
\label{eq:nstar}
\end{equation}
Assuming a monochromatic PBH mass spectrum, Eq.~\eqref{eq:spect0} can be written as
\begin{equation}
\frac{dn_a}{dk}(k_{\ast},\tau_{\rm BH})\simeq \frac{f_{\rm ev}\kappa_a}{4\pi^2}\left(\frac{m_P}{M}\right)^2T_{\rm BH}^2\mathcal{I}\left(\frac{k_{\ast}}{T_{\rm BH}}\right)\, ,
\label{eq:alpdens}
\end{equation}
where $T_{\rm BH}$ is given by Eq.~\eqref{eq:bh_temp} and is taken constant, and
\begin{equation}
\mathcal{I}(x)=
x^2\int_0^1 \frac{\theta^{-2}(1-\theta)^{2/3}}{\exp\left[x\theta^{-2/3}(1-\theta)^{1/3}\right]-1}\,d\theta\, ,
\label{eq:F}
\end{equation}
where we have performed the change of variable $\theta=t/\tau_{\rm BH}$, and
$x=k_{\ast}/T_{\rm BH}$. This expression is similar to those obtained in Ref.~\cite{Auffinger:2020afu}. Notice however that in using Eq.~\eqref{eq:at} we have assumed that the transition from the radiation dominated to the matter dominated era is instantaneous. We have solved Eq.~\eqref{eq:FRW}  numerically for $M$ in the range $10^1$~g~$\leq M\leq 10^9$~g we have shown that the $a_{\ast}$ differs for a factor $0.97$ from those calculated in Eq.~\eqref{eq:astar}. To obtain a better approximation Eq.~\eqref{eq:alpdens} should thus be corrected by a factor $\alpha_c=1.09$. Figure \ref{fig:comp_vs_approx} shows a comparison between the present-day scalar Hawking spectra from a population of PBHs of mass $M=10^5\,{\rm g}$ calculated both using the exact cosmological evolution as well as an instantaneous transition to matter domination.

 \subsection{Redshift to the present epoch}
 
As a further step, Eq.~\eqref{eq:alpdens} has to be redshifted to the present epoch, in order to be expressed as a function of the present-day quantities. Introducing the (standard) redshift parameter $1+z_\ast=a(t_0)/a(\tau_{\rm BH})$ and recalling that $dn_a/dk\sim a^{-2}$, one obtains
 \begin{eqnarray}
\label{eq:spectral_density4}
 	\frac{dn_a}{dk}(k,t_0)&\simeq& 3.27\times10^{-3}\left(\frac{m_P}{M}\right)^2
	T_{{\rm BH},0}^2\mathcal{I}\left(\frac{k}{T_{{\rm BH},0}}\right)	\nonumber \\
	&\simeq& 2.02\times 10^{-5}\left(\frac{10^5{\rm g}}{M}\right)^2\left(\frac{T_{{\rm BH},0}}{1{\rm eV}}\right)^2\mathcal{I}\left(\frac{k}{T_{{\rm BH},0}}\right)\,{\rm cm}^{-3}\,{\rm keV}^{-1}
	\,,
 \end{eqnarray}
where $k$ is the present-day ALP momentum, and we have defined $T_{{\rm BH},0}=T_{\rm BH}/(1+z_\ast)$ and thus used the fact that $k_\ast/T_{\rm BH}=k/T_{\rm{BH},0}$. 
To evaluate $z_\ast$, consider that at $\tau_{\rm BH}$ the Universe exits the matter dominated epoch (where the Hubble parameter $H=2/3t$) and becomes radiation dominated again ($\rho=\rho_r$). The cosmic expansion rate is given by the Friedmann equation as $H^2 = 8 \pi G \rho/3$.
  From the latter equation we find
 \begin{equation}
 \label{eq:radiation_density}
 	\rho_r(\tau_{\rm BH})=\frac{m_P^2}{6\pi\tau_{\rm BH}^2}=\frac{\pi^2}{30}g_\ast(T_\ast)T_\ast^4\,\ ,
 \end{equation}
where $g_\ast(T)$ is the number of relativistic degrees of freedom contributing to the radiation energy density (see e.g. \cite{kolb}), given by a good approximation by
\begin{equation}
\label{eq:energy_dof}
g_\ast(T_\ast)=\sum_B g_B\left(\frac{T_B}{T_\ast}\right)^4+\frac{7}{8}\sum_F g_F\left(\frac{T_F}{T_\ast}\right)^4 \,\ .
\end{equation} 
In Eq.~\eqref{eq:energy_dof} the sums run over all bosons $B$ (fermions $F$), having $g_B$ ($g_F$) intrinsic degrees of freedom, temperature $T_B$ ($T_F$) and rest mass below $T_\ast$.~\footnote{A better evaluation of  degrees of freedom can be found in \cite{Husdal:2016haj}.}
Inverting Eq.~\eqref{eq:radiation_density} and using Eq.~\eqref{eq:bh_time} one finds
 \begin{equation}
 \label{eq:T_star}
	T_\ast \simeq 890\left(\frac{100}{g(T_\ast)}\right)^{1/4}\left(\frac{10^{5} \,\textrm{g}}{M}\right)^{3/2}\,\textrm{GeV}\,.
 \end{equation}

Furthermore we consider the entropy density~\cite{kolb} 
\begin{equation}
	s=\frac{2\pi^2}{45}g_S(T_\ast) T_\ast^3 \,\ ,
\end{equation}
 where $g_S(T)$ are the entropic degrees of freedom, given by
 \begin{equation}
\label{eq:entropy_dof}
	g_S(T_\ast)=\sum_B g_B\left(\frac{T_B}{T_\ast}\right)^3+\frac{7}{8}\sum_F g_F\left(\frac{T_F}{T_\ast}\right)^3\,.
\end{equation}
  Note that the various components of cosmic radiation may not be at equilibrium with photons due to thermal decoupling, and thus 
  $g_\ast(T)$ and $g_S(T)$  generally have different values. However, at the high temperatures of the early Universe, one may assume that all particles of the Standard Model are relativistic and at thermal equilibrium with each other, and thus set $g_\ast(T_\ast) = g_S(T_\ast)\simeq 100$. 
  Since entropy conservation implies that the quantity $a^3g_S(T)T^3$ remains constant, imposing that
 \begin{equation}
 \label{eq:entropy_cons}
 	a_\ast^3g_S(T_\ast)T_\ast^3=a_0^3g_S(T_0)T_0^3 \,\ ,
 \end{equation}
 one finds
 \begin{equation}
 	1+z_\ast=\frac{a_0}{a_\ast}=\left(\frac{g_S(T_\ast)}{g_S(T_0)}\right)^{1/3}\frac{T_\ast}{T_0}\,,
 \end{equation}
 where $a_\ast=a(\tau_{\rm BH})$, $T_0={2.73} \, \textrm{K}=2.35 \times 10^{-4} \,\textrm{eV}$~\cite{Zyla:2020zbs} is the current temperature of the cosmic microwave background (CMB) and $g_S(T_0)=3.91$~\cite{kolb} is the corresponding number of entropy degrees of freedom. Using these values, as well as Eqs.~\eqref{eq:bh_temp} and \eqref{eq:T_star}, we obtain
 \begin{equation}
 \label{eq:T_BH0}
 	T_{{\rm BH},0}=\frac{T_{\rm BH}}{1+z_\ast}\simeq 9.47\left(\frac{100}{g_S(T_\ast)}\right)^{1/12}
	\left(\frac{M}{10^{5}\,\textrm{g}}\right)^{1/2}\,\textrm{eV}\,,
 \end{equation}
which represents the black hole temperature redshifted to the present epoch. 

\begin{figure}
	\centering
	\includegraphics[width=.8\textwidth]{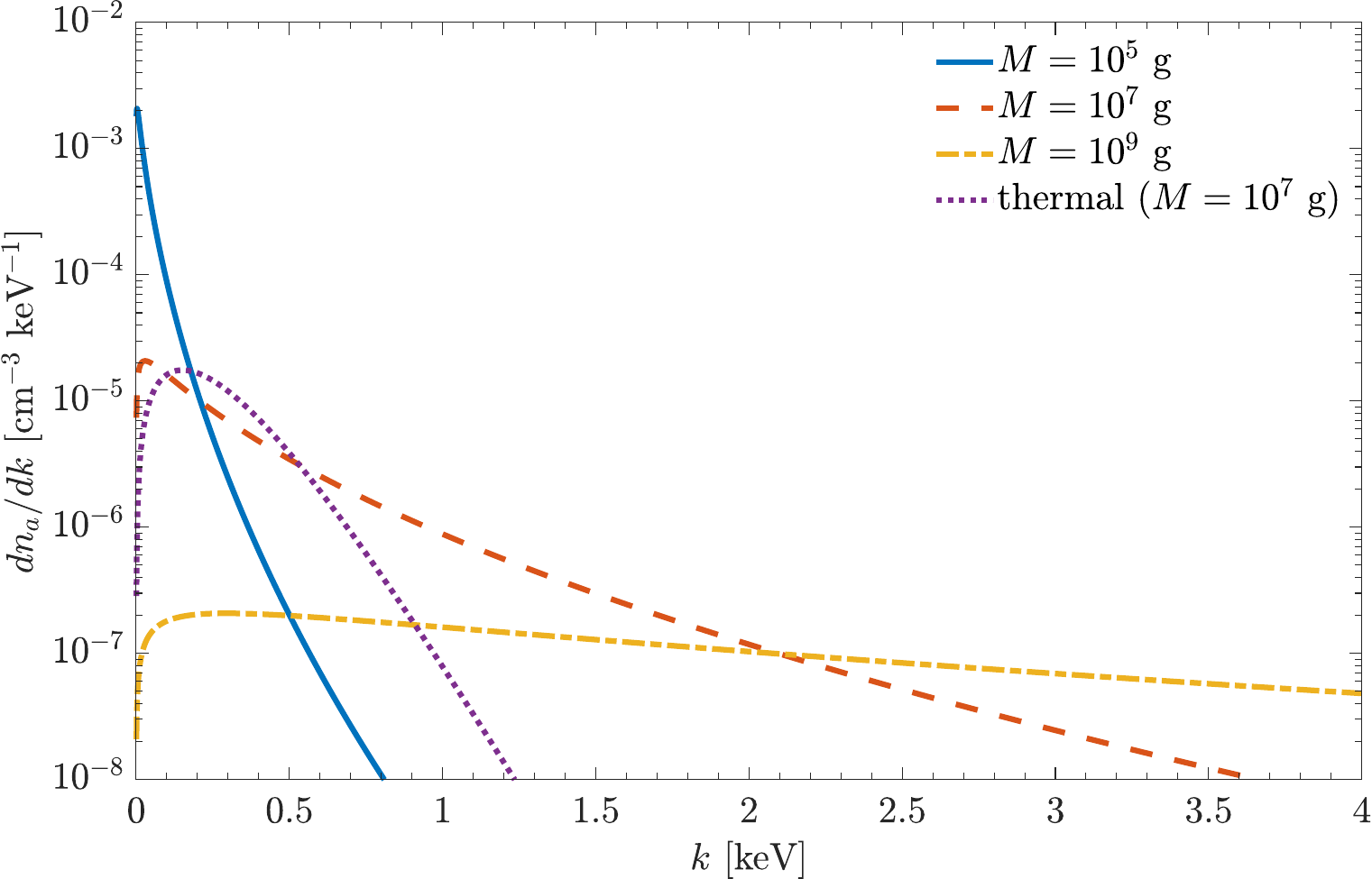}
		\caption{ALP number density  energy spectrum  for three different values of the PBH mass $M$. A purely thermal spectrum at a temperature corresponding to $M = 10^7\,{\rm g}$ is also shown for comparison.
		}
	\label{fig:spectra}
\end{figure}

In Figure~\ref{fig:spectra} we plot the ALP number density spectrum obtained from Eq.~\eqref{eq:spectral_density4}
for different PBH masses.
We realize that the ALP spectrum extends in the soft X-rays range. In particular, increasing the mass of the PBH one finds spectra with longer
tails extending towards high energies.
Indeed, from Eq.~\eqref{eq:T_BH0} one realizes that increasing the PBH mass also $T_{\rm{BH}, 0}$ increases consequently. {Finally, we note that for a given temperature $T_{\rm BH,0}$ the Hawking spectrum peaks at a lower energy and exhibits a longer high-energy tail with respect to the corresponding thermal blackbody spectrum.}

The total ALP number density is obtained by integrating  Eq.~\eqref{eq:spectral_density4}, which gives
 \begin{eqnarray}
 \label{eq:particle_density}
 	n_a= \int_0^\infty \frac{dn_a}{dk}(k,t_0)d k 
 	&=&1.18\times10^{-2} \left(\frac{m_P}{M}\right)^2 T_{{\rm BH},0}^3 \nonumber	\\
 	&=&6.18\times10^{-5}\left(\frac{100}{g_S(T_\ast)}\right)^{1/4}
	\left(\frac{10^{5}\,\textrm{g}}{M} \right)^{1/2}\,\textrm{cm}^{-3}\,,
 \end{eqnarray}
where we have used the relation $\int_0^\infty\mathcal{I}(x)dx=3\zeta(3) \simeq 3.60617$, $\zeta$ being the Riemann zeta function.

 \subsection{ALP energy density spectrum and extra-radiation}

Let us focus for the moment on very light ALPs, $m_a\ll 1$~keV. In this case ALPs are still relativistic, $k\simeq \omega$ where $\omega$ is the ALP energy. From Eq.~\eqref{eq:spectral_density4} we can obtain the energy density of ALPs per unit of energy   
 \begin{eqnarray}
 \label{eq:spectrum_1}
 	\frac{d \rho_a}{d \omega}&=&\omega\frac{dn_a}{dk}(\omega,t_0)=3.27\times10^{-3}
	\left(\frac{m_P}{M}\right)^2T_{{\rm BH},0}^3\left(\frac{\omega}{T_{{\rm BH},0}}\right)\mathcal{I}\left(\frac{\omega}{T_{{\rm BH},0}}\right)	 \nonumber \\
	&=&2.02\times10^{-8}\left(\frac{10^5{\rm g}}{M}\right)^2\left(\frac{T_{{\rm BH},0}}{1{\rm eV}}\right)^3\left(\frac{\omega}{T_{{\rm BH},0}}\right)\mathcal{I}\left(\frac{\omega}{T_{{\rm BH},0}}\right)\,{\rm cm}^{-3}
	\,\ .
 \end{eqnarray}
We also compute the total energy density as
 \begin{eqnarray}
 \label{eq:energy_density_tot}
 	\rho_a =\int_0^\infty{d \omega} \omega\frac{dn_a}{dk}(\omega,t_0)
 	&=& 5.14\times 10^{-2} \left(\frac{m_P}{M} \right)^2T_{{\rm BH},0}^4 \nonumber 	\\
	&=& 2.49\times 10^{-3} \left(\frac{100}{g_S(T_\ast)}\right)^{1/3}\,\textrm{eV}\, \textrm{cm}^{-3}\,,\end{eqnarray}
where we have used $\int_0^\infty x\mathcal{I}(x)dx=4\pi^5/45\sqrt{3}\simeq 15.7$.
The average energy per ALP is thus
\begin{equation}
 \label{eq:mean_energy}
\varepsilon_a=\rho_a/n_a=40.3\,\left(\frac{100}{g_S(T_\ast)}\right)^{1/12}
\left(\frac{M}{10^{5}\,\textrm{g}}\right)^{1/2}\,\textrm{eV}\,.
\end{equation}

This ALP background might behave as dark radiation for 
$m_a \lesssim 10$~MeV (see~\cite{Hooper:2019gtx}), if they are stable on cosmological time scales (see Sec.~6),
contributing to an extra-component of the effective number of neutrinos
\begin{equation}
\label{eq:DNeff}
	\Delta N_{\rm eff}=\frac{8}{7}\left(\frac{11}{4}\right)^{4/3}\frac{\rho_{a}(T_0)}{\rho_{\rm CMB}(T_0)}\,\ .
\end{equation}
 The present CMB density is $\rho_{\rm CMB}(T_0)=0.26$~eV$/$cm$^3$~\cite{kolb}. Using Eq.~\eqref{eq:energy_density_tot} we obtain 
\begin{equation}
	\Delta N_{\rm eff}\simeq 0.042\left(\frac{100}{g_S(T_*)}\right)^{1/3}\,,
\end{equation}
This value is of the same order of magnitude of the $\Delta N_{\nu,{\rm eff}}=0.046$ due to partial reheating of neutrinos~\cite{Mangano:2005cc}.~\footnote{Refined calculations \cite{Akita:2020szl,Bennett:2020zkv} have recently modified this value to $\Delta N_{\nu,{\rm eff}}=0.044$, but the difference is not relevant for the datasets considered in this work.}

The latest  2018 data from the Planck satellite experiment give $N_{\rm eff}=2.99 \pm 0.17$~\cite{Aghanim:2018eyx}. 
Therefore, the ALP extra-radiation would be compatible with the current data.
Future experiment CMB-S4 would reach a sensitivity to $\Delta N_{\rm eff} \leq 0.06$  
at $95\%$~\cite{Baumann:2015rya} confidence level, which shows that probing this scenario is a challenging task in the near future.


\section{ALP-photon conversions}

\subsection{Conversions in homogeneous magnetic field}

Signatures of ultralight ALPs (with $m_a \lesssim 10^{-9}$~eV) produced by PBH are associated to  conversions into photons in the cosmic
magnetic fields. 
Therefore, we devote this Section to present the formalism necessary to characterize these phenomena.

The ALP-photon interaction is described by the following  Lagrangian~\cite{DiLuzio:2020wdo}
\begin{equation}
\label{mr}
{\cal L}_{a\gamma}=-\frac{1}{4} \,g_{a\gamma}
F_{\mu\nu}\tilde{F}^{\mu\nu}a=g_{a\gamma} \, {\bf E}\cdot{\bf B}\,a~,
\end{equation}
where $g_{a\gamma}$ is the photon-ALP coupling constant (which has the dimension of an inverse energy),
$a$ is the ALP field, 
$F_{\mu\nu}$ is the electromagnetic field and $\tilde{F}_{\mu\nu}$ is its dual.
This interaction
allows for photon-ALP mixing in the presence of an external magnetic field ${\bf B}$~\cite{Raffelt:1987im,sikivie,Anselm:1987vj}.

In order to describe the basics of this phenomenon at first we consider the simple case in which we have 
a photon beam
of energy $\omega$ propagating
along the $x_3$ direction in a cold ionized and magnetized medium with 
an homogeneous magnetic field ${\bf B}$.  We indicate with ${\bf B}_T$ the transverse component of the  magnetic field,  in the plane normal to the beam direction and we  choose the $x_2$-axis along ${\bf B}_T$ so that $B_1$ vanishes.
We indicate 
 the linear photon polarization states parallel  and orthogonal to  the transverse field direction ${\bf B}_T$, as   $A_{\parallel}$ and $A_{\perp}$,
 respectively. 
 In this situation,  the linearized equations of motion for the ALP-photon system are~\cite{Raffelt:1987im}
\begin{equation}
\label{vne}
\left(\omega- i \frac{\partial}{\partial x_3} + {\mathcal M} \right)
 \left(\begin{array}{c} A_{\perp} \\ A_{\parallel} \\a
 \end{array}\right)
= 0 \,\ ,
\end{equation}
where the mixing matrix can be written as~\cite{Mirizzi:2005ng,Mirizzi:2006zy}
\begin{equation}
{\cal M} = \left(\begin{array}{ccc}
\Delta_{ \perp}   & 0 & 0 \\
0 &  \Delta_{ \parallel}   & \Delta_{a \gamma}  \\
0 & \Delta_{a \gamma} & \Delta_a
\end{array}\right)~,
\label{eq:massgen}
\end{equation}
whose elements are~\cite{Raffelt:1987im} $\Delta_\perp \equiv \Delta_{\rm pl} + \Delta_{\perp}^{\rm CM} + \Delta_{\textrm{CMB}},$ $ \Delta_\parallel \equiv \Delta_{\rm pl} + \Delta_{\parallel}^{\rm CM} + \Delta_{\textrm{CMB}},$ $\Delta_{a\gamma} \equiv {g_{a\gamma} B_T}/{2} $ and $\Delta_a \equiv - {m_a^2}/{2\omega}$, where $m_a$ is the ALP mass. The term $\Delta_{\rm pl} \equiv -{\omega^2_{\rm pl}}/{2\omega}$ takes into account plasma effects, where $\omega_{\rm pl}$ is the plasma frequency. This can be  expressed as a function of the electron density in the medium $n_e$ as $\omega_{\rm pl} \simeq 3.69 \times 10^{- 11} \sqrt{n_e /{\rm cm}^{- 3}} \, {\rm eV}$. The  terms $\Delta_{\parallel,\perp}^{\rm CM}$ describe the Cotton-Mouton  effect, i.e.~the birefringence of fluids in the presence of a transverse magnetic field.  A vacuum Cotton-Mouton effect is expected from QED one-loop corrections to the photon polarization in the presence of an external magnetic field $\Delta_\mathrm{QED} = |\Delta_{\perp}^{\rm CM}- \Delta_{\parallel}^{\rm CM}| \propto B^2_T$, but this effect is completely negligible in the present context. 
The term $\Delta_{\textrm{CMB}} \propto \rho_{\textrm{CMB}}$ represents the contribution to the photon polarization 
by the CMB radiation \cite{Dobrynina:2014qba}. 
 One numerically finds
\begin{eqnarray}  
\Delta_{a\gamma}&\simeq &   1.52\times10^{-3} \left(\frac{g_{a\gamma}}{10^{-12}\textrm{GeV}^{-1}} \right)
\left(\frac{B_T}{10^{-9}\,\rm G}\right) {\rm Mpc}^{-1}
\nonumber\,,\\  
\Delta_a &\simeq &
 -7.8 \times 10^{5} \left(\frac{m_a}{10^{-10} 
        {\rm eV}}\right)^2 \left(\frac{\omega}{{\rm keV}} \right)^{-1} {\rm Mpc}^{-1}
\nonumber\,,\\  
\Delta_{\rm pl}&\simeq & 
  -1.1\times10^{-2}\left(\frac{\omega}{{\rm keV}}\right)^{-1}
         \left(\frac{n_e}{10^{-7} \,{\rm cm}^{-3}}\right) {\rm Mpc}^{-1}
\nonumber\,,\\
\Delta_{\rm QED}&\simeq & 
4.1\times10^{-18}\left(\frac{\omega}{{\rm keV}}\right)
\left(\frac{B_T}{10^{-9}\,\rm G}\right)^2 {\rm Mpc}^{-1} \nonumber\,,\\
\Delta_{\textrm{CMB}}&\simeq & 2.62\times 10^4 \left(\frac{T}{1\,\textrm{eV}}\right)^4\left(\frac{\omega}{{\rm keV}} \right)  {\rm Mpc}^{-1}
 \,\ .
\label{eq:Delta0}\end{eqnarray}

For the above estimates that we will use in the following as benchmark values, we refer to the following physical inputs: The strength of $B$-fields and the electron density $n_e$ in the previous equations are typical values for the intergalactic medium~\cite{Mirizzi:2006zy}.
Concerning the ALP-photon coupling we take it below the bound obtained from ultralight ALPs ($m_a \lesssim 10^{-10}$~eV)
from the absence of $\gamma$-rays from SN~1987A~\cite{Grifols:1996id,Brockway:1996yr}. 
At this regard a recent analysis constrains $g_{a\gamma} \lesssim 5.3 \times 10^{-12}$~GeV$^{-1}$ for 
$m_a \lesssim 4.4 \times 10^{-10}$~eV~\cite{Payez:2014xsa}.
We note that in the equations only the product $g_{a\gamma} B_T$ enters, thus we will often present our bounds on this parameter.
In particular, 
when we will discuss  possible ALP-photon conversions during the recombination epoch in the early Universe,
we require $g_{a\gamma} B_T \lesssim 10^{-13}$~GeV$^{-1}$~nG for $m_a \lesssim 10^{-9}$~eV in order to avoid an excessive distortion of the CMB spectrum \cite{Mirizzi:2009nq}  caused by the
conversions during that phase. 

Concerning the ALP energy today the bulk of the spectrum is in the (sub)-keV energy range, the higher the PBH mass the broader
the tail of the spectrum (see Fig. \ref{fig:spectra}).
Finally, we will assume $m_a \ll \omega_{\rm pl}$, neglecting the term $\Delta_a$ in the evolution, since a  value of $m_a$ much larger than
$ \omega_{\rm pl}$
would suppress the conversions. 

In the case of a homogeneous magnetic field, the ALP-photon mixing described by Eq.~\eqref{eq:massgen} reduces to a $2\times2$ problem involving only
the $(A_{\parallel}, a)$ fields.
If we define the conversion probability at $x_3$ as the probability of finding a photon after starting with a pure ALP beam, we may express it as
\begin{equation}
\label{eq:2x2_prob_ag}
	P_{a\gamma}(x_3)=|A_\parallel(x_3)|^2
	=\sin^2(2\theta)\sin^2\left(\frac{\Delta_{\rm osc} x_3}{2}\right)
	=\sin^2(2\theta)\sin^2\left(\frac{\pi x_3}{\ell_{\rm osc}}\right)\,,
\end{equation}
where
\begin{equation}
\tan 2 \theta= \frac{2 \Delta_{a\gamma}}{\Delta_\parallel-\Delta_a} \,\ ,
\end{equation}
\begin{equation}
	\Delta_{\rm osc}=\sqrt{(\Delta_\parallel-\Delta_a)^2+4\Delta_{a\gamma}^2} \,\ ,
	\label{eq:deltaosc}
\end{equation}
and $\ell_{\rm osc}=2\pi/\Delta_{\rm osc}$ is the ALP-photon oscillation length. In the approximation of small mixing angle, $\theta\ll 1$ or $\Delta_{a\gamma}\ll\Delta_\parallel$, and large oscillation length, $\Delta_{\rm osc}\ll 1$, the previous expression simply reduces to
\begin{equation}
	\label{eq:2x2_prog_ag_approx}
	P_{a\gamma}(x_3)\simeq(\Delta_{a\gamma}x_3)^2\,.
\end{equation}
In order to gain more physical intuition, it is convenient to rewrite Eq.~\eqref{eq:2x2_prob_ag} as
\begin{equation}
P_{a\gamma}^{(0)}=\left( \Delta_{a\gamma} x_3 \right)^2 \textrm{sinc}^2\left( \Delta_{\rm osc} x_3 /2 \right) \,\ ,
\label{eq:prob0}
\end{equation}
where the sinc function is defined as $\textrm{sinc}\,x \equiv x^{-1}\sin x$.
We now notice that, for the conditions we are interested in, 
the $\Delta_{a \gamma}^2$ term in Eq.~\eqref{eq:deltaosc}  is always irrelevant.

\subsection{Conversions in the early Universe}

We can extend the previous treatment in the case relevant for us of ALP-photon conversions in the early Universe.
This is a topic that has been treated in a series of papers (see, e.g.~\cite{Christensson:2002ig,Mirizzi:2009nq}).
In particular, here we follow the treatment given in~\cite{Evoli:2016zhj} to which we address the interested reader for further details.

\subsubsection{Magnetic domains}

Assuming the existence of  primordial $B$-fields, Planck data constrain its amplitude 
to be less than a few nG coherent on a scale $l \sim {\mathcal O}$(1 Mpc)~\cite{Ade:2015cva}.
Therefore, we will model the primordial $B$-field as a network of cells with size set by its coherence length
(see also~\cite{Jedamzik:2018itu} for a recent study). 
 The strength of ${\bf B}$ is assumed to be the same in every domain, but  its direction changes in a random way  from one cell to another. 
Because of this, $A_{\parallel}$ in one cell is not the same as $A_{\parallel}$ in the next one. 
Therefore the propagation over many magnetic domains represents  a full 3-dimensional case.

\subsubsection{Redshift}

 As the particle beam propagates across magnetic domains, one must also account for the expansion of the Universe, represented by the scale factor $a$ or equivalently by the redshift parameter $1+z=a_0/a(t)$. The relationship between cosmological time and redshift is given by 
\begin{equation}
\label{eq:redshift-vs-time}
	\frac{d z}{d t}=-(1+z)H_0\sqrt{\Omega_{\Lambda}+\Omega_{m}(1+z)^3}\,,
\end{equation}
where $H_0\simeq {67.4}\, \textrm{km} \,\textrm{s}^{-1}\,\textrm{Mpc}^{-1}$, $\Omega_{\Lambda}\simeq 0.685$ and
 $\Omega_{m}\simeq 0.315$~\cite{Zyla:2020zbs} are the present-day values of the Hubble parameter, the dark energy density parameter and the matter density parameter respectively. 

At an epoch characterized by redshift $z$, the length of a magnetic domain is reduced as
\begin{equation}
	l(z)=\frac{l_0}{1+z}
\end{equation}
and the beam takes a time $\Delta t(z)=l(z)$ (in natural units) to cross it. Therefore, one can construct a sequence of redshift values as follows: starting from a large value of the redshift parameter $z_0$, each value of the sequence is defined by 
	\begin{equation}
	\label{eq:redshift_sequence}
		z_{n+1}=z_n+\Delta z_n\,,
	\end{equation}
where $\Delta z_n<0$ represents the redshift variation corresponding to the time interval $\Delta t_n = l_n = l_0/(1+z_n)$ the beam takes to cross the $n$-th cell. 

The values of the matrix coefficients defined by Eqs.~\eqref{eq:Delta0} in the $n$-th cell are thus related to their present-day values by
\begin{align}
		\Delta_{a}^{(n)} &= {\Delta_{a}^{(0)}}({1+z_n})^{-1}		\\
		\Delta_{a\gamma}^{(n)} &= \Delta_{a\gamma}^{(0)}(1+z_n)^2	\\
		\Delta_{\rm QED}^{(n)} &= \Delta_{\rm QED}^{(0)}(1+z_n)^5	\\
		\Delta_{\rm pl}^{(n)} &= \Delta_{\rm pl}^{(0)}(1+z_n)^2	\\
		\Delta_{\rm CMB}^{(n)} &= \Delta_{\rm CMB}^{(0)}(1+z_n)^5\,,
		\label{eq:matrix2}
	\end{align}
	where we recall that electron number density, magnetic field\footnote{By the conservation of magnetic flux.}, energy and temperature are redshifted as $n_e(z)=n_{e0}(1+z)^3$, $B(z)=B_0(1+z)^2$, $\omega(z)=\omega(1+z)$ and ${T(z)=T_0(1+z)}$ respectively. By comparing these values,
	 as noted in~\cite{Evoli:2016zhj}, one finds that for $z\gtrsim 2 \times 10^3$ the mixing is strongly suppressed due to the CMB term dominating over all other coefficients. We may therefore choose $z_0=2 \times 10^3$ and study conversions only for $z<z_0$. We may also expect strong \emph{resonant} conversion effects when 
	$|\Delta_{\rm QED}+\Delta_{\rm pl}+\Delta_{\rm CMB}|=0$, as in this case the mixing term $\Delta_{a\gamma}$  would dominate. These effects should correspond to large peaks in the conversion probability, when expressed as a function of $z$.

\begin{figure}
 	\centering
 	\includegraphics[width=.8\textwidth]{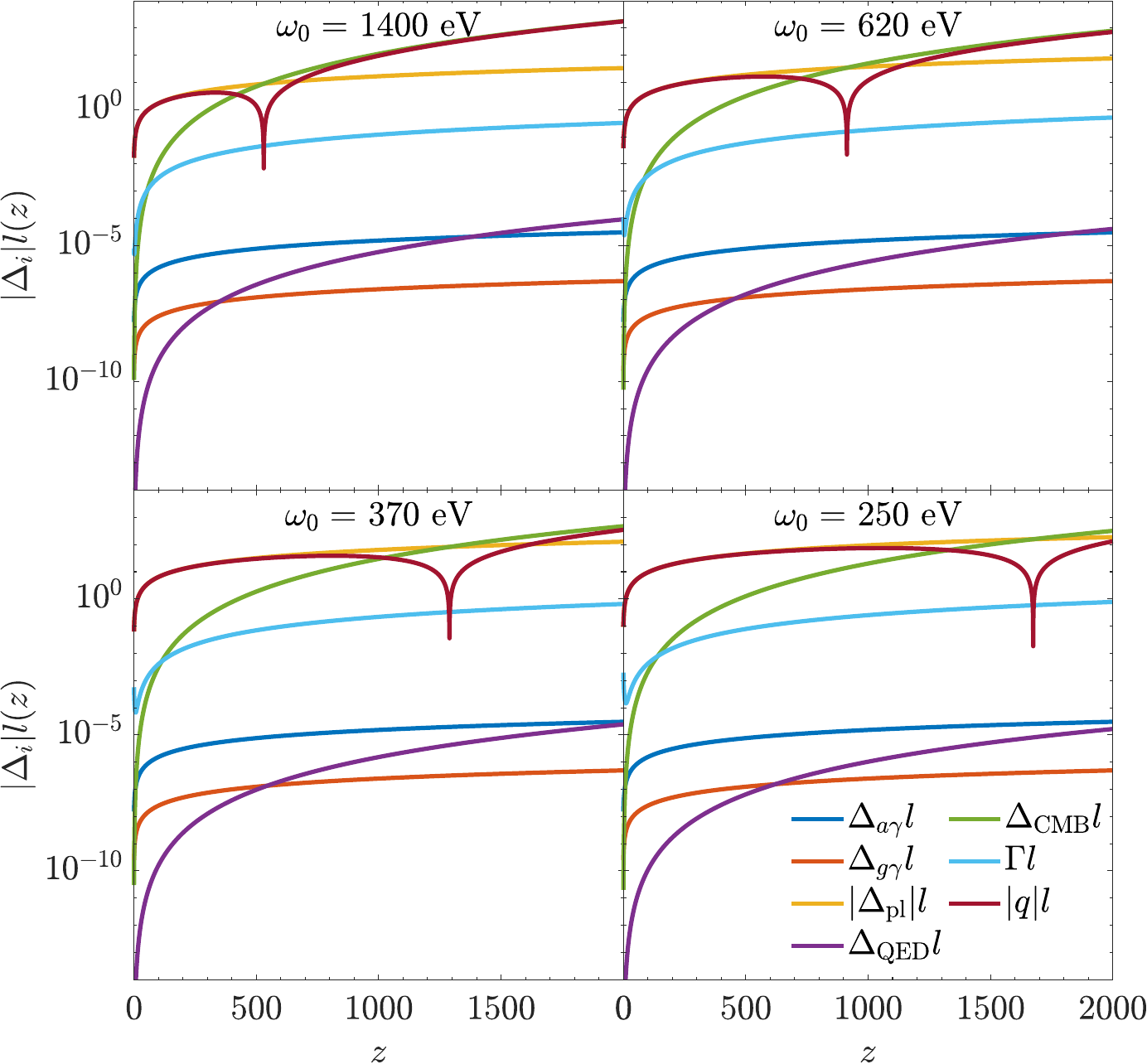}
 \caption{Plots of the matrix elements in Eqs.~\eqref{eq:matrix2}, as well as $|q|=|\Delta_{\rm QED}+\Delta_{\rm pl}+\Delta_{\rm CMB}|$ and the
  $\Gamma$ coefficient defined in Eq.~\eqref{eq:gamma}, 
  multiplied by domain length $l(z)$ and expressed as functions of redshift $z$ for several values of the present-day beam energy $\omega$. The fiducial values $g_{a\gamma}=10^{-17} \,\textrm{GeV}^{-1}$ and $B_{T 0}={1}\,\textrm{nG}$ have been used.}
 	\label{fig:matrix_coefficients}
 \end{figure}

The quantities in Eqs.~\eqref{eq:matrix2} are plotted in Fig.~\ref{fig:matrix_coefficients}, for several values of the present-day energy $\omega$. We have assumed that $m_a$ is small enough so that $\Delta_a$ becomes negligible at high redshift, and we have used the same fiducial values
 as in~\cite{Evoli:2016zhj}, namely $g_{a\gamma}=10^{-17} \,\textrm{GeV}^{-1}$ and $B_{T 0}={1}\,\textrm{nG}$.

\subsubsection{Photon absorption}

As discussed above, ALP conversions into photons start 
at $z \sim 2 \times 10^3$, and thus take place in the phase of matter-radiation decoupling started at $z \sim 1100$~\cite{kolb}.
During this epoch, at $T \sim 0.3$ eV, the process $H + \gamma \leftrightarrow p + \text{e}^-$ goes out of equilibrium, 
leading to the \emph{recombination} of electrons and protons into Hydrogen and Helium.
After this event the Universe entered the so-called ``Dark Ages'', during which virtually all H and He atoms were neutral (except for a very low fraction of ionized hydrogen, $X_e\simeq 10^{-4}$) and no luminous sources were present. This period was later followed, at 
$z\lesssim 12$~\cite{Adam:2016hgk}, by a significant \emph{reionization} of the intergalactic medium (to be discussed in Sec.~5).

Then, during the ALP-photon conversions we have to take into account  the absorption of photons by hydrogen and helium atoms formed at recombination. 
The photon absorption on electrons and atoms adds a damping term in the ALP-photon Hamiltonian~\cite{Mirizzi:2009aj,Kartavtsev:2016doq}
\begin{equation}
{\mathcal M} \to {\mathcal M} + \textrm{diag}\left(- i \frac{\Gamma}{2}, -i \frac{\Gamma}{2}, 0\right) \,\ .
\end{equation}
In the absorption term
one has to take into account the photo-electric and to the Compton effect.
 Thus we express the total photon absorption rate as~\cite{Evoli:2016zhj}
\begin{equation}
\label{eq:gamma}
	\Gamma(\omega)=\sigma_{\rm H}^{\rm PE}(\omega) n_{\rm H}+\sigma_{\rm He}^{\rm PE}(\omega) n_{\rm He}+\sigma_{\rm KN}(\omega)n_e\,,
\end{equation}
where $\sigma_{\rm H}^{\rm PE}$ and $\sigma_{\rm He}^{\rm PE}$ are the photoelectric cross-sections for Hydrogen and Helium. The symbols $n_{\rm H}$ and $n_{\rm He}$ represent the number densities of H and He nuclei,~\footnote{We remind that, after recombination and before reionization, virtually every nucleus corresponds to a neutral atom, as the ionization fraction is very small.} whose present-day values are related to the helium fraction $Y_p$ and to the baryon number density $n_{b0}$ as
\begin{align}
\label{eq:H_He_density}
	n_{\rm H0}=(1-Y_p)n_{b0}\,,	&&	n_{\rm He0}=\frac{Y_p}{4} n_{b0}\,,
\end{align}
while $n_e$ is the total number density of electrons (both free and bound in atoms). The cross-sections for the photoelectric effect may be evaluated from the analytic fits provided in~\cite{Verner:1996th}. The (integrated) Klein-Nishina cross-section for the Compton effect is instead given by~\cite{greiner2009}. We neglect the pair production process which is negligible at the energies we are interested in.

\subsubsection{Recursive expression for photon-ALP conversions}

In~\cite{Evoli:2016zhj} it has been shown that under the condition 
\begin{equation}
\Delta_{a \gamma} \ll \Delta_{\rm pl} \,\ ,
\end{equation}
which for us is satisfied independently of the redshift (see Fig.~\ref{fig:matrix_coefficients}),
one  obtains an average conversion probability in the $n$-th magnetic domain
\begin{equation}
P^{(n+1)}_{a\gamma} = \left[P^{(n)}_{a\gamma} + 
P^{0 (n)}_{a\gamma} \right] \text{e}^{-\Gamma_n l_n} \,\ ,
\label{eq:pconvpert}
\end{equation}
where the subscript $n$ refer to the values of the different quantities at redshift $z_n$, and 
$P^{0}_{a\gamma}$ is given by Eq.~\eqref{eq:prob0}.

\begin{figure}
\centering
\includegraphics[width=.8\textwidth]{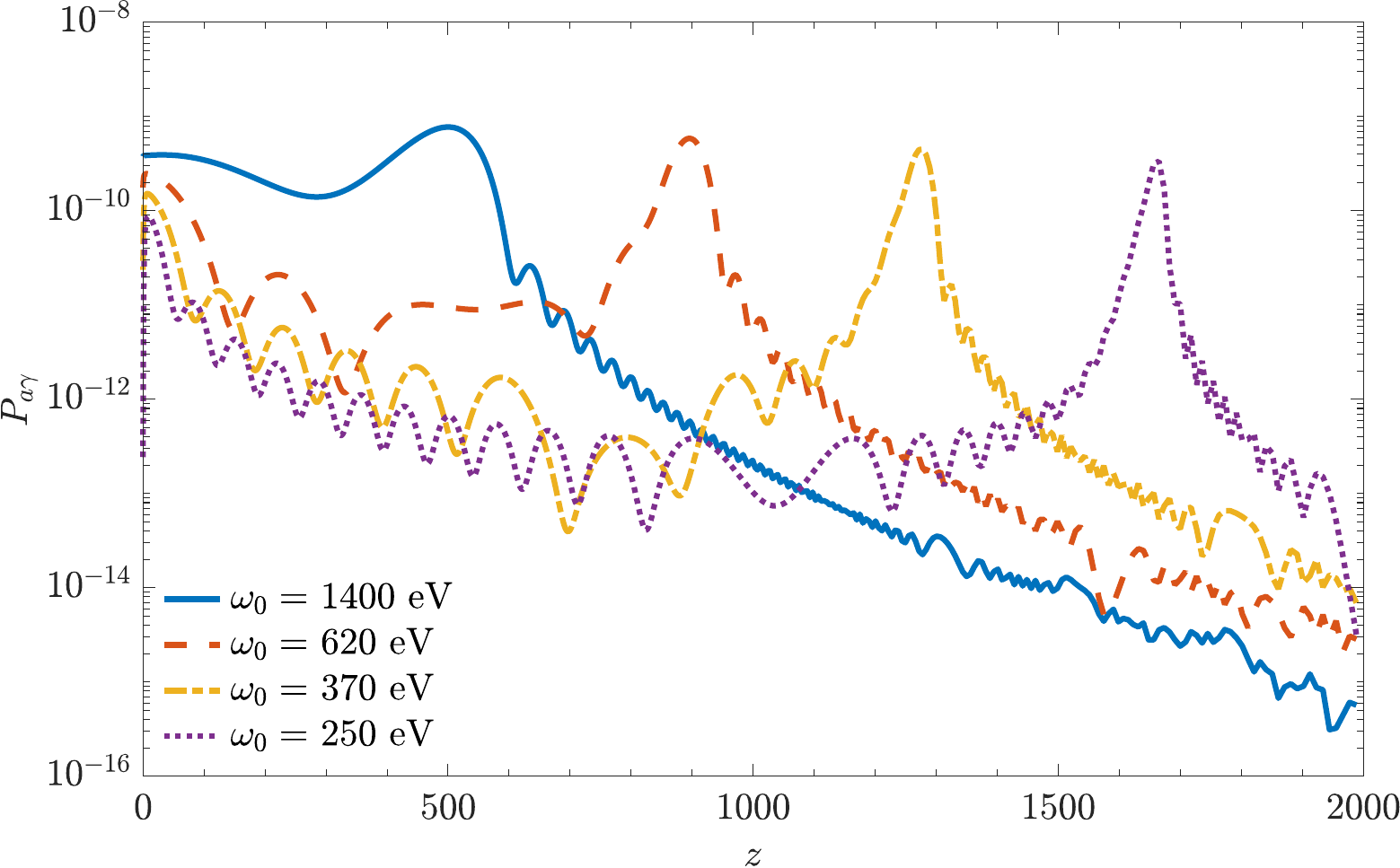}
\caption{ALP-photon conversion probabilitiy as a function of redshift $z$ for several values of the present-day beam energy $\omega$,
for  $g_{a\gamma}=10^{-17}\,\textrm{GeV}^{-1}$ and $B_{T 0}={1} \,\textrm{nG}$.}
\label{fig:conv_prob}
\end{figure}

The ALP-photon conversion probability is shown in Fig.~\ref{fig:conv_prob} for several values of the present-day beam energy $\omega$, taking
 $g_{a\gamma}=10^{-17}\,\textrm{GeV}^{-1}$ and $B_{T 0}={1} \,\textrm{nG}$ as fiducial values. Comparing these results to those in 
 Fig.~\ref{fig:matrix_coefficients}, we note that peaks in the conversion probability correspond to the values of $z$ at which
  $|\Delta_{\rm QED}+\Delta_{\rm pl}+\Delta_{\rm CMB}|=0$, thus confirming the resonant effects discussed above.

\section{X-ray fluxes from  ultra-light ALP conversions}

Equipped with the results obtained in the previous Section we are now ready to calculate the conversions of the ALP flux, 
produced by PBH evaporation, into X-rays.
 We present here bounds and possible signatures
that can be obtained from different observables in the case of ultralight ALPs ($m_a \lesssim 10^{-9}$~eV). {Indeed, we checked that the photon-ALP conversion probability in the intergalactic magnetic field is significantly suppressed for
 $m_a \gtrsim 10^{-9}$~eV. Therefore, the bounds reported below and in the next Section refer to ALP masses smaller than this threshold.}

\subsection{Cosmic X-ray background}

The diffuse 
\textit{cosmic X-ray background} (CXB), first observed in 1962~\cite{Giacconi:1962zz}, covers photon energies in the range $\sim 0.1$--${100}\,\textrm{keV}$. Since its discovery, a number of space-based observatories have studied the CXB, in the attempt to determine its origin. At present, it has been possible to resolve most of the CXB in the $0.5$--$10$~keV band into discrete sources, many of which have been found to be active galactic nuclei (AGNs); for a review, see~\cite{Brandt:2015cja}. The unresolved cosmic X-ray background has been measured by a series of experiments and can be  represented by different analytical fits (see~\cite{Ballesteros:2019exr}).

In this context, we explore the possibility that ALPs emitted by primordial black holes in the early Universe are converted into photons which contribute to the still-unresolved part of the CXB. Notice from Fig.~\ref{fig:spectra} that the energies of these particles range between a few keV and tens of keV; Therefore, the photons arising by conversions would indeed produce a signal in the X-ray spectrum.

\begin{figure}
	\centering
	\includegraphics[width=.8\textwidth]{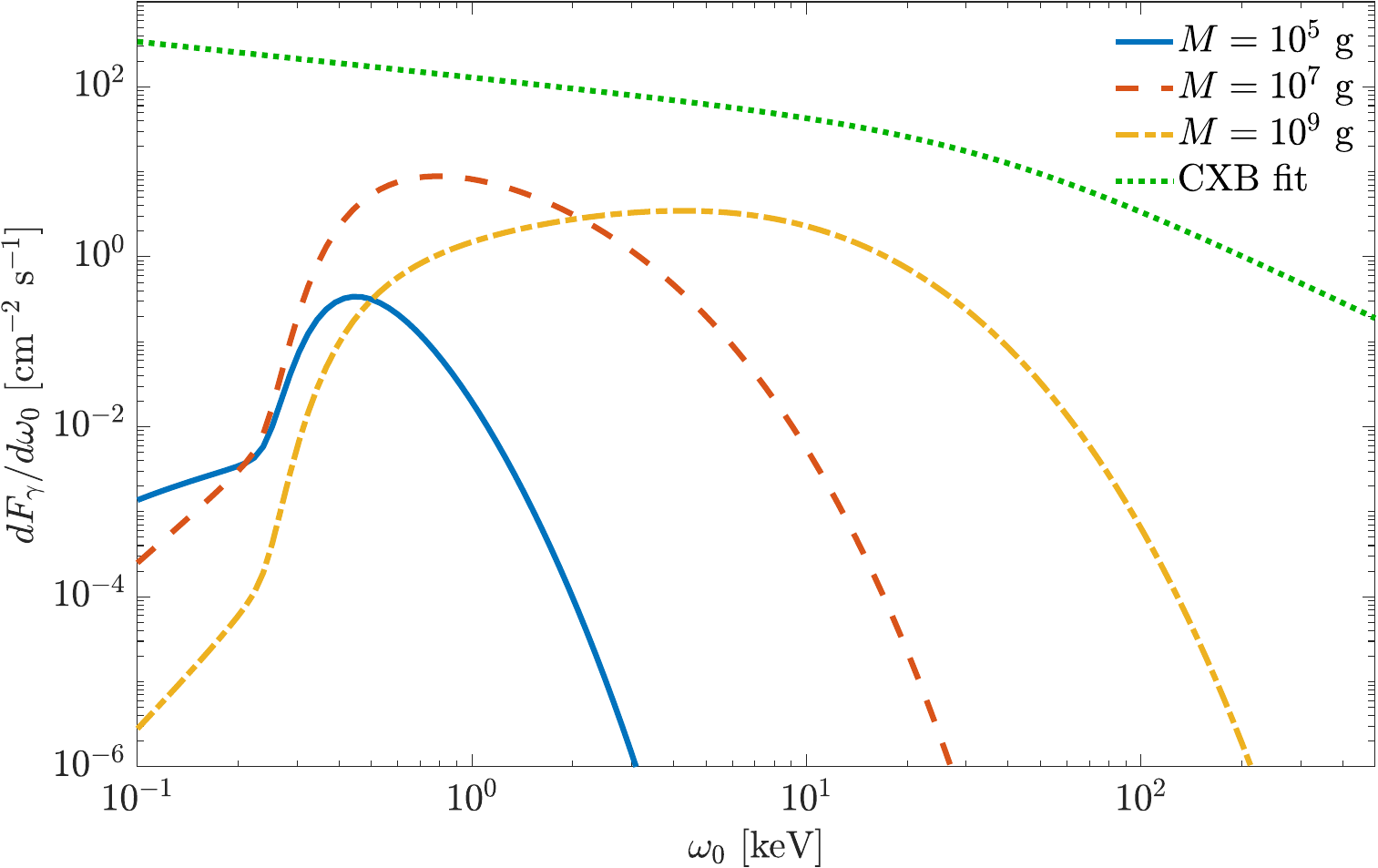}
	\caption{Present-day photon fluxes from ALP conversions for different values of the PBH mass $M$ with 
	$g_{a\gamma} B_{T0}=10^{-14}\,\textrm{GeV}^{-1}\,\textrm{nG}$, plotted along with the double power-law fit of the CXB flux data from Ref. \cite{Ballesteros:2019exr}.}
	\label{fig:photon_flux}
\end{figure}

The photon flux from  ALP conversions during the evolution of the Universe is then given by
\begin{equation}
\label{eq:photon_flux}
	\frac{d F_\gamma}{d \omega} (\omega,t_0)=\frac{d F_a}{d \omega} (\omega,t_0)P_{a\gamma}(\omega,t_0) \,\ ,
	\end{equation}
where the present-day conversion probability $P_{a\gamma}(\omega,t_0)$ is  obtained by iterating Eq.~\eqref{eq:pconvpert} up to the present epoch with a present-day beam energy $\omega$. 
The ALP \textit{spectral energy flux} can be obtained by rewriting Eq.~\eqref{eq:spectral_density4} in appropriate units
as
\begin{equation}
\label{eq:particle_flux}
	\frac{d F_a}{d \omega} (\omega,t_0)\simeq 6.06\times10^{2}\left(\frac{10^5{\rm g}}{M}\right)^2\left(\frac{T_{{\rm BH},0}}{1{\rm eV}}\right)^3\left(\frac{\omega}{T_{{\rm BH},0}}\right)\mathcal{I}\left(\frac{\omega}{T_{{\rm BH},0}}\right)\, \,\, \textrm{cm}^{-2}\,\textrm{s}^{-1}\,.
\end{equation}
Photon fluxes from ALP-photon conversions are shown in Fig.~\ref{fig:photon_flux} for different values of the PBH mass $M$
and $g_{a\gamma} B_{T0}=10^{-14}\, \textrm{GeV}^{-1}\,\textrm{nG}$. We note that the photon fluxes from heavier PBHs are displaced towards higher energies.

\begin{figure}
	\centering
	\includegraphics[width=.82\textwidth]{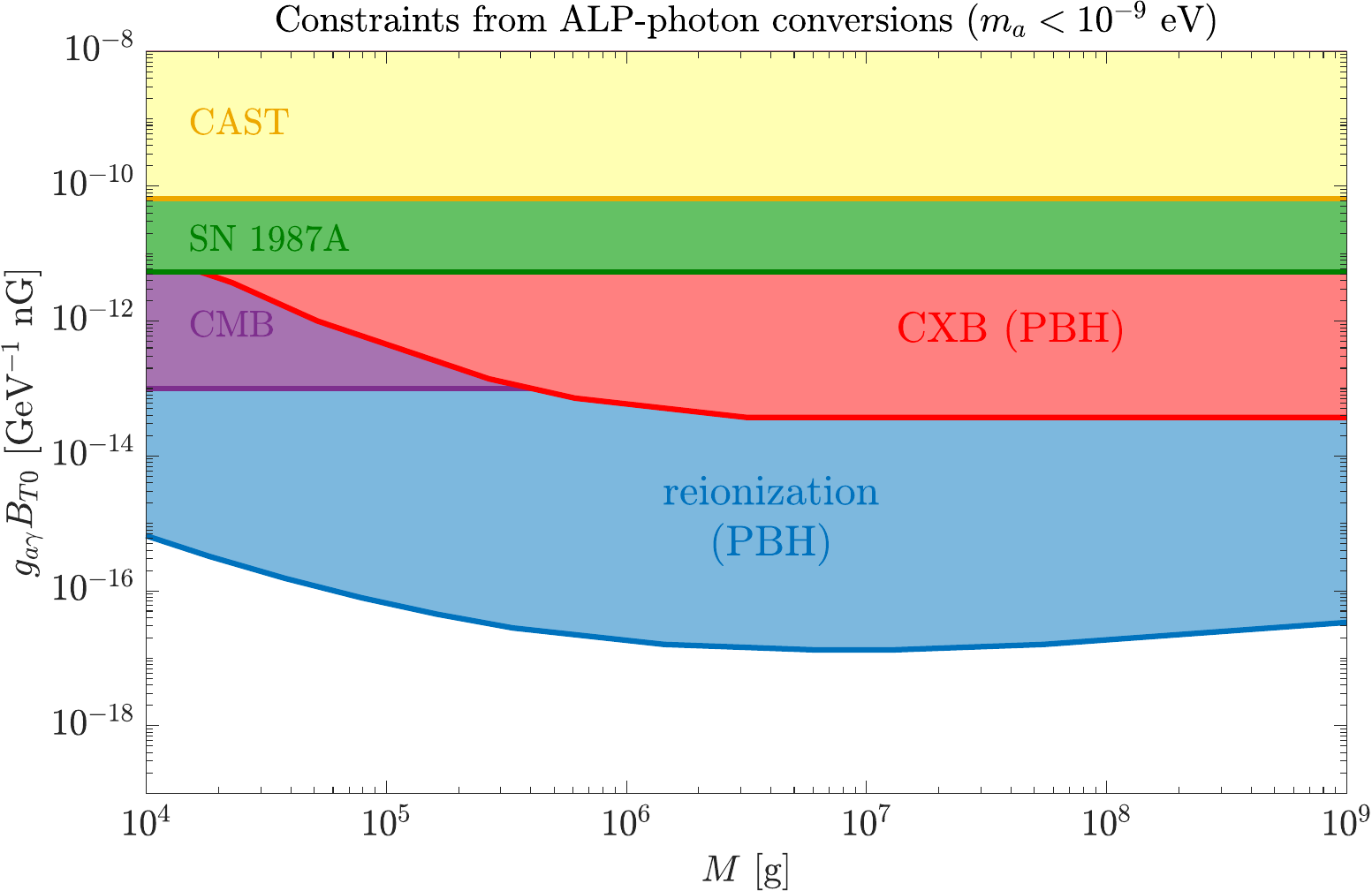}
	\caption{Values of $g_{a\gamma} B_{T0}$ excluded both by CXB and reionization for each value of the PBH mass $M$, shown as shaded areas in the parameter space. The previous bounds  on $g_{a\gamma}$ from the CAST search on solar ALPs~\cite{Anastassopoulos:2017ftl}, from the absence of gamma-rays from conversions of ALPs emitted by SN 1987A~\cite{Payez:2014xsa} and from CMB distortions~\cite{Mirizzi:2005ng} are shown for comparison, multiplied by a present-day magnetic field $B_{T0}=1$~{nG}. {Note that the photon-ALP conversion probability in the intergalactic magnetic field is significantly suppressed for
 $m_a \gtrsim 10^{-9}$~eV. Therefore, our bounds refer to smaller ALP masses.}
	}
	\label{fig:conversion_bounds}
\end{figure}

In order to compare these results with experimental data, we use the double power-law fit provided in Ref. \cite{Ballesteros:2019exr}, assuming it extends down to $0.1\,{\rm keV}$. In order to  constrain the  ALP-photon mixing parameter, we impose that the photon fluxes  from ALP conversions do not exceed this fit curve. The region of the parameter space excluded as a result of this comparison is shown as a shaded area in Fig.~\ref{fig:conversion_bounds}. One realizes that for $M \simeq 10^9$~g the strongest bound is obtained, namely {$g_{a\gamma} B_{T0} \lesssim 10^{-14} \,\textrm{GeV}^{-1}\,\textrm{nG}$}. This bound worsens for lower PBH masses, eventually becoming not competitive with other constraints for $M \simeq 10^4$~g. Our constraints are compared to those from the CAST experiment on solar ALPs~\cite{Anastassopoulos:2017ftl}, from the absence of gamma-rays from ALP conversions emitted by SN 1987A~\cite{Payez:2014xsa} and from the CMB distortions induced by ALP conversions~\cite{Mirizzi:2005ng}.

Although the CXB bound seems to be dominated by the one from reionization, we must remark that the latter depends upon the ambient magnetic fields during ALP conversion. In particular, although possible existence of a primordial magnetic field of cosmological origin has been the subject of an intense investigation during the last few decades~\cite{Durrer:2013pga}, there is no evidence for it, and only upper limits are reported. Regarding intergalactic magnetic field, an upper limit of $1.7$~nG has been recently derived~\cite{Pshirkov:2015tua}, although there is also a lower limit of about $3\times10^{-7}$~nG reported in~\cite{Neronov:2010gir} which stems from the non-observation of GeV gamma-ray emission from electromagnetic cascade initiated by TeV gamma rays in intergalactic medium. Since the strength of primordial and intergalactic magnetic fields are uncertain,
it is not redundant to present constraints based on different assumptions concerning the cosmic magnetic fields.

\subsection{Soft X-ray excess in Galaxy Clusters}

Starting with the observations of the Virgo and Coma clusters by the EUVE and ROSAT space telescopes in 1996~\cite{lieu,lieub,bowyer}, an excess was found in the ``soft'' part of the X-ray spectrum (at sub-keV energies). This feature was later confirmed by the study of several other Galaxy Clusters (see e.g.~\cite{Bonamente:2002wh,Durret:2008jn} for reviews), but its origin is still unknown.

In Ref.~\cite{Conlon:2013txa} it was suggested that the soft X-ray excess could be due to conversions of ALPs of cosmological origin into photons, made possible by the magnetic fields present in the intracluster medium. The origin of this cosmic axion background (CAB) was assumed to be the decay of moduli after inflation in string cosmology. For the soft excess in the Coma cluster, this hypothesis was later explored with much greater detail
 in~\cite{Angus:2013sua}, where it was favored over other possible soft X-ray production mechanisms such as emission by a warm gas ($\sim {0.1}$~{keV}) or inverse Compton scattering in the intracluster medium.

Here,  following the estimation of  Ref.~\cite{Conlon:2013txa} we  show that the evaporating PBH in our scenario could equally well be a source of cosmic ALPs responsible for the soft X-ray excess in the Coma cluster or other similar objects.
In the specific case of the Coma cluster the  X-ray excess luminosity is given by~\cite{Conlon:2013txa} 
\begin{equation}
\mathcal{L}\simeq {1.6 \times 10^{42}}\, \textrm{erg} \,\textrm{s}^{-1} \,\ .
\end{equation}
 We start by estimating the energy density of the ALP background that from 
 Eq.~\eqref{eq:energy_density_tot} can be written as
\begin{equation}
\label{eq:axion_density}
	\rho_a\simeq 1.18\times10^{59}
	\left(\frac{100}{g_S(T_*)}\right)^{1/3}\,\textrm{erg}\,\textrm{Mpc}^{-3}\,.
\end{equation}
This value is about one order of magnitude below that predicted in~\cite{Conlon:2013txa}. We take the ALP-photon conversion probability in the approximation of small mixing angle and large oscillation length, given by Eq.~\eqref{eq:2x2_prog_ag_approx}, which we rewrite as
\begin{equation}
\label{eq:prob_xcess}
	P_{a\gamma}\simeq	\frac{g_{a\gamma}^2}{4}\langle B_T^2\rangle L^2\simeq 4.61\times 10^{-8}
	\left(\frac{g_{a\gamma}}{10^{-13}\,\textrm{GeV}^{-1}} \right)^2
	\left(\frac{B}{{2}\, \upmu{\rm G}}\right)^2
	\left(\frac{L}{{1}\,\textrm{kpc}}\right)^2\,.
\end{equation}
Here $\langle B_T^2\rangle\simeq B^2/2$, where $B$ is the intracluster magnetic field and $\langle B_T^2\rangle$ its average component transverse to the beam trajectory, and $L$ is the distance travelled by the beam. The average intensity magnetic field in the Coma cluster is $B\simeq {2}\,\  \upmu{\rm G}$, with a correlation length in the 2{--}{34} {kpc} range~\cite{Bonafede:2010xg}.  Multiplying Eq.~\eqref{eq:axion_density} by Eq.~\eqref{eq:prob_xcess} (and dividing by $L$) we find that the photon luminosity per $\textrm{Mpc}^3$ given by
\begin{equation}
	\mathcal{L}_{\textrm{Mpc}^3}=5.28\times 10^{40}  
	\left(\frac{100}{g_S(T_*)}\right)^{1/3}
	\left(\frac{g_{a\gamma}}{10^{-13} \,\textrm{GeV}^{-1}}\right)^2
	\left(\frac{B}{{2} \, \upmu{\rm G}}\right)^2
	\left(\frac{L}{{1}\,\textrm{kpc}}\right)
	\,\textrm{erg} \,\textrm{Mpc}^{-3}\,\textrm{s}^{-1}
\end{equation}
that, multiplied by a cylindrical volume with radius ${0.5}\,\textrm{Mpc}$ and length ${3}\,\textrm{Mpc}$, gives
\begin{equation}
	\mathcal{L}=1.24\times 10^{41} 
	\left(\frac{100}{g_S(T_\ast)}\right)^{1/3}
	\left(\frac{g_{a\gamma}}{10^{-13}\,\textrm{GeV}^{-1}}\right)^2
	\left(\frac{B}{{2}\,  \upmu{\rm G}}\right)^2
	\left(\frac{L}{{1}\,\textrm{kpc}}\right)
	\,\textrm{erg}\,\textrm{s}^{-1}\,.
\end{equation}
This result is just about one order of magnitude below that given in Ref.~\cite{Conlon:2013txa}, and it shows that for $g_{a\gamma} 
\gtrsim 10^{-13}\,\textrm{GeV}^{-1}$, the cosmic ALP background from PBH evaporation could be responsible for a substantial part, if not all, of the soft X-ray excess in the Coma cluster.

\section{Effects of ALP conversions on reionization}

\begin{figure}
\centering
	\includegraphics[width=.8\textwidth]{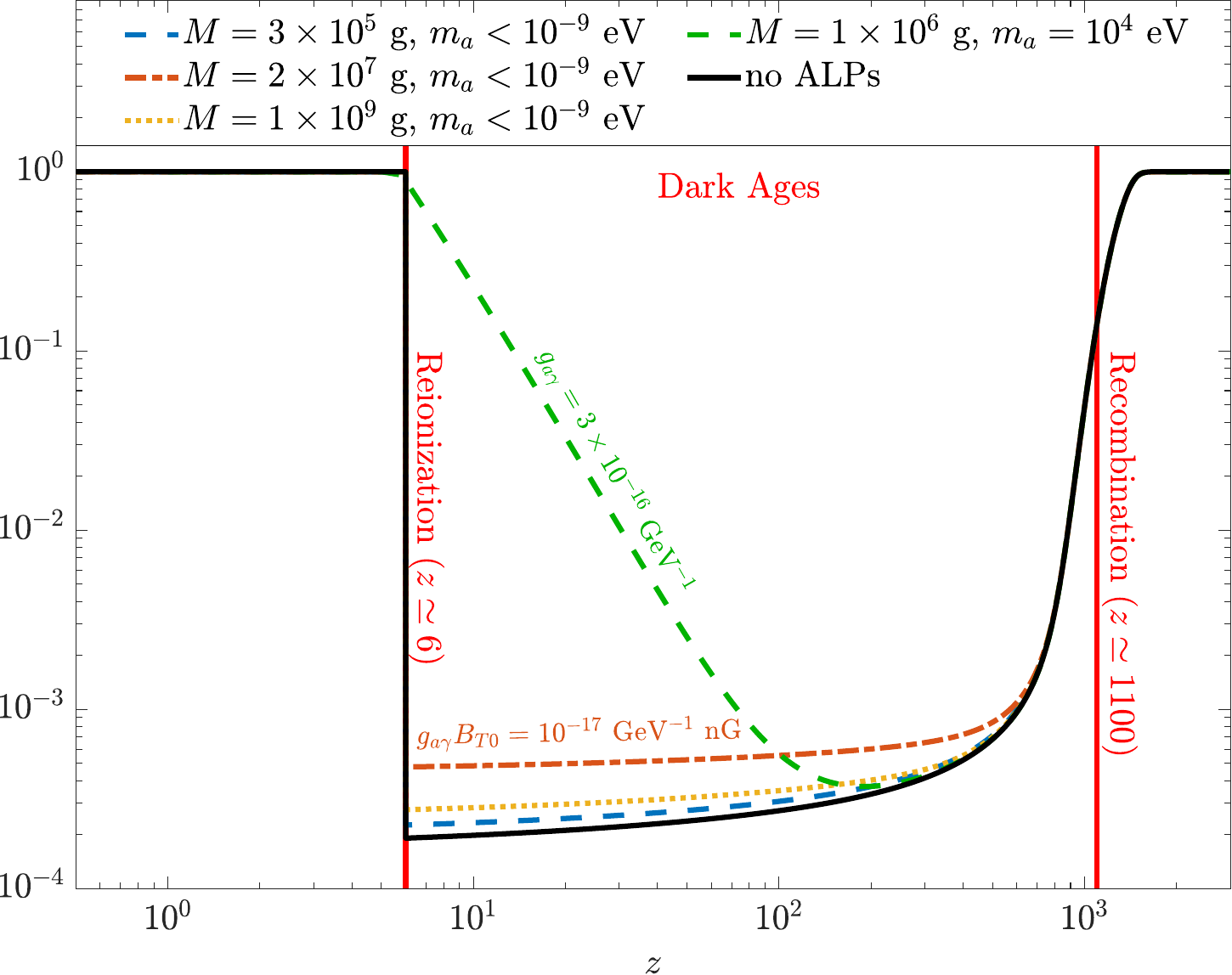}
	\caption{Ionization fraction as a function of redshift during the cosmic Dark Ages, both in the standard scenario and including contributions from light ALP-photon conversions and from massive ALP decays. Different values of the mass $M$ of PBHs producing the ALP flux have been tested. 
	The values $g_{a\gamma}=
	10^{-17} \,\textrm{GeV}^{-1}$ and $B_{T0}=\textrm{nG}$ have been used for the ultralight ALP case, whereas the massive case 
	($m_a=10^4$~eV) for decaying ALPs
	was computed taking $g_{a\gamma}=3\times10^{-16}\,{\rm GeV}^{-1}$.}
	\label{fig:ionization_alp}
\end{figure}

In order to assess the impact of ALP conversions on cosmological reionization we closely follow the approach 
of Ref.~\cite{Evoli:2016zhj}.
In Fig.~\ref{fig:ionization_alp} we show with a continuous curve the evolution of the ionization fraction
$X_{\rm e}$ after recombination in the standard scenario (as computed with the RECFAST code~\cite{recfast}), 
defined as \cite{kolb}
\begin{equation}
\label{eq:X_e}
	X_e=\frac{n_p}{n_b}=\frac{n_e^{\rm free}}{n_b}\,,
\end{equation}
where $n_p$, $n_e^{\rm free}$ and $n_b$ are the number densities of free protons, free electrons and baryons, respectively.
We have  assumed that reionization occurred instantaneously at $z = 6$.
The high-frequency photons ($\omega\in [10,\, 10^3$]~{eV}) produced by ALP conversions in the primordial magnetic fields 
 could have helped ionize the intergalactic medium during the Dark Ages after recombination, and thus could have had some observable effects on the process of reionization. We now turn our attention to these effects.

We start by estimating the contribution of free electrons produced in the $n$-th cell (at redshift $z_n$) by ionizing photons deriving from conversions of ALPs. The spectrum of photons produced by conversions at redshift $z_n$ is given by
\begin{equation}
\label{eq:photon_spectrum}
	\frac{d n_\gamma}{d \omega_n}=\frac{d n_a}{d \omega} (1+z_n)^2P_{a\gamma}^{(n)}(\omega)\,,
\end{equation}
where ${d n_a}/{d \omega} $ is the present-day ALP Hawking spectra,  as given by Eq.~\eqref{eq:spectrum_1}, and each is multiplied by a factor $(1+z_n)^2$ to account for the expansion of the Universe. The ALP-photon conversion probability in the $n$-th cell $P_{a\gamma}^{(n)}$ is
 given by Eq.~\eqref{eq:pconvpert}.

 The number of photons not yet absorbed by the intergalactic medium in the $n$-th cell is given by
\begin{equation}
	f_S(z_n)=1-e^{-\Gamma_n l_n}\,,
\end{equation}
where $\Gamma$ is the absorption rate given by Eq.~\eqref{eq:gamma}, $l$ is the domain length, and as usual the index $n$ denotes the values of these quantities in the $n$-th cell. Furthermore, the photon energy fraction contributing to ionization (as opposed to heat dissipation, atomic excitations, etc.) is given by~\cite{Evoli:2012zz}
\begin{equation}
	f_I(z_n)=a \left[1-\left(X_e(z_n)\right)^b\right]^c\,,
\end{equation}
where $a=0.3846$, $b=0.5420$ and $c=1.1952$. We find that the number density of free electrons produced in the $n$-th cell by photons coming from ALP conversions is given by
\begin{equation}
\label{eq:D_free_elec}
	\Delta n_e^{\rm free}(z_n)=\int_0^\infty d{\omega_n} \frac{d n_\gamma}{d \omega_n}f_S(z_n)f_I(z_n)\frac{\omega_n}{{13.6}\,\textrm{eV}}\,,
\end{equation}
where $ d n_\gamma/d \omega_n$ is given by Eq.~\ref{eq:photon_spectrum}, $\omega_n/{13.6}\,\textrm{eV}$ is the approximate number of hydrogen atoms that can be ionized by a photon of energy $\omega_n$. The ionization fraction $X_e(z_n)$ is therefore obtained by adding to the standard value $X_e^0(z_n)$ all the contributions due to cells characterized by a redshift parameter greater than $z_n$, i.e.
\begin{equation}
	X_e(z_n)=X_e^0(z_n)+\sum_{i=1}^n\frac{\Delta n_e^{\rm free}(z_i)}{n_{\rm H}(z_i)}\,,
\end{equation}
where $n_{\rm H}(z_i)=n_{\rm H0}(1+z_i)^3$ is the hydrogen density at redshift $z_i$ and $n_{\rm H0}$ is its current value as given by Eq. \eqref{eq:H_He_density}. We have also taken into account the recombination of free electrons with ionized hydrogen~\cite{Cirelli:2009bb}, although we have verified that this contribution is always negligible.

\begin{figure}
	\centering
	\includegraphics[width=.8\textwidth]{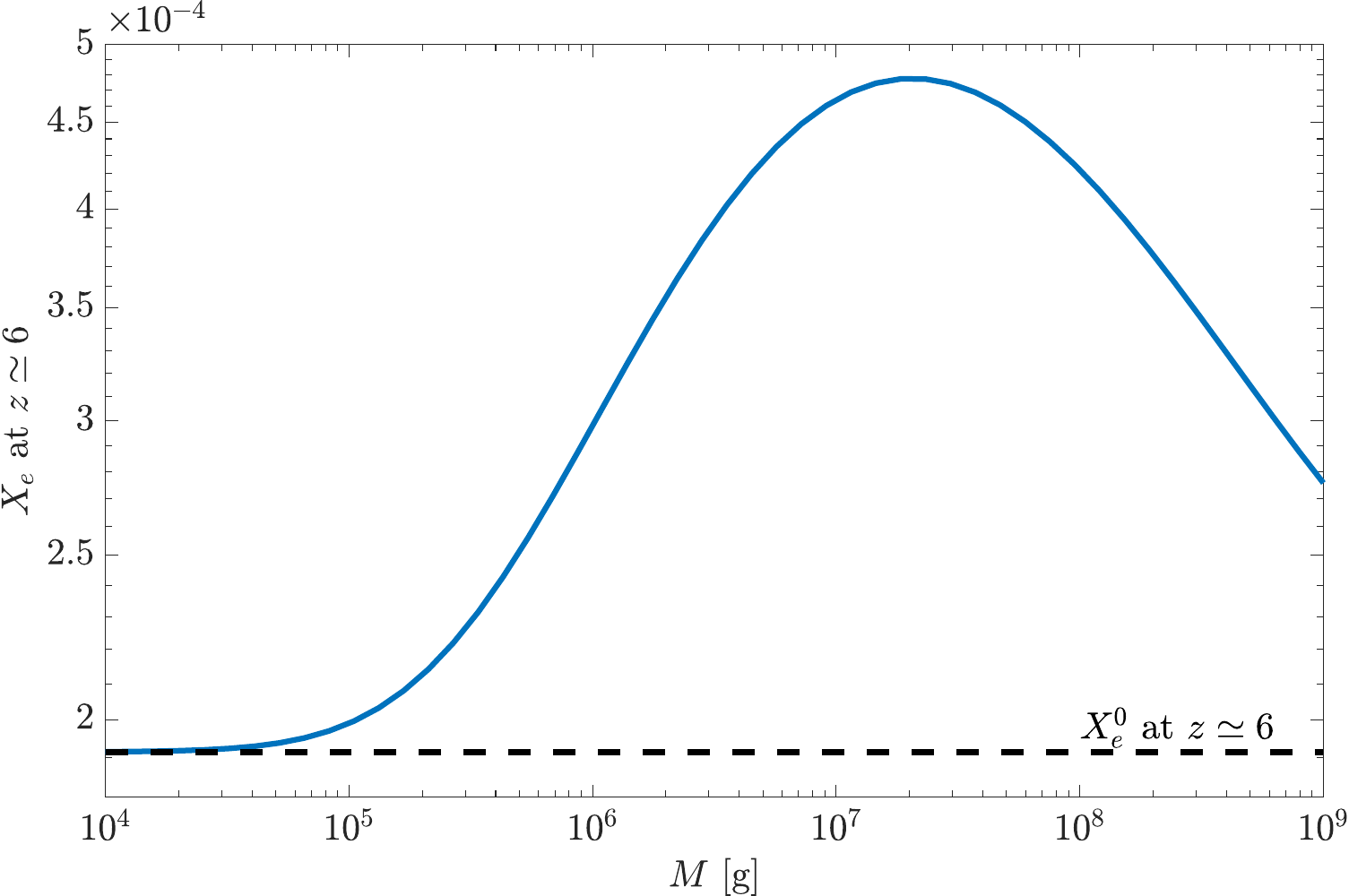}
	\caption{Ionization fraction just before reionization, at $z\simeq 6$, as a function of the PBH mass. The dashed line represents the standard value in the absence of ALP-photon conversions. The value  $g_{a\gamma}B_{T0}=
	10^{-17} \,\textrm{GeV}^{-1}\,\textrm{nG}$ has been used for the ALP-photon mixing parameter.}
	\label{fig:ionization-vs-mass}
\end{figure}

This procedure has been carried out numerically for different values of the PBH mass $M$. The results are compared to the standard scenario in 
Fig.~\ref{fig:ionization_alp}, where the value $g_{a\gamma}B_{T0}=
	10^{-17} \,\textrm{GeV}^{-1}\,\textrm{nG}$  has been used for the ALP-photon mixing parameter. We have verified that effects of ALP-photon conversions on reionization are appreciable only for a PBH mass range  $ 10^{4} \,\textrm{g}\lesssim M\lesssim 10^{9} \,\textrm{g}$.
Moreover, ionization is not ``monotonic'' with respect to the PBH mass. To see this explicitly, in Fig.~\ref{fig:ionization-vs-mass} we have plotted the ionization fraction just before reionization, at $z\simeq 6$, as a function of the PBH mass. It can be seen that the impact of ALP-photon conversions is negligible for $M\lesssim 10^{4} \,\textrm{g}$, while it is maximum for $M\simeq 1.84 \times 10^{7} \,\textrm{g}$. Intuitively, this can be understood by looking at Fig.~\ref{fig:spectra}: for smaller $M$, the Hawking spectrum is peaked at lower energies, and thus most particles do not possess enough energy to reionize the IGM; on the other hand, for larger $M$, the spectrum gets so spread out at high energies that there are too few
particles per unit energy to produce significant reionization. Thus, there is only a relatively narrow mass range in which PBH evaporation has a noticeable impact on reionization.

To directly compare our results with experimental measurements, we estimate the value of the Thomson optical depth of the Universe in our scenario, defined as
\begin{equation}
\label{eq:opt_depth}
	\tau=\int_0^\infty d{z}\left |\frac {dt}{dz}\right|\sigma_T n_e^{\rm free}(z)\,,
\end{equation}
where $n_e^{\rm free}(z_n)$ is the free electron density at redshift $z_n$ and $\sigma_T\simeq 6.65 \times 10^{-25}\,\textrm{cm}^2$ is the Thomson cross-section. The most recent experimental determination of this quantity is 
\begin{equation}
\label{eq:tau_exp}
	\tau_{\rm exp}=0.054\pm 0.007\,,
\end{equation}
as obtained by the 2018 results of the Planck collaboration~\cite{Adam:2016hgk}.

 In the case of magnetic domains, Eq.~\eqref{eq:opt_depth} can be approximated as
\begin{equation}
	\tau\simeq\sum_n l_n \sigma_T n_e^{\rm free}(z_n)\,,
\end{equation}
where the sum extends over all domains crossed by the beam, and we have used the fact that in the $n$-th cell $d t\simeq \Delta t_n = l_n$. In order to impose conservative bounds, following~\cite{Evoli:2016zhj}, we assume that \emph{all} ionization of the intergalactic medium at $z>6$ is due to the ALP production and conversion mechanism we described. We thus conservatively ignore all stars and other sources of ionizing radiation that may have contributed to the density of free electrons during this epoch. On the other hand,  we assume that for $z\leq 6$ the intergalactic medium is completely ionized; in this case, the density of free electrons can simply be obtained by redshifting its current value, $n_{e0}$. In summary, we model the free electron density at redshift $z_n$ as
\begin{equation}
\label{eq:free_elec}
	n_e^{\rm free}(z_n)=
	\begin{cases}
		n_{e0}(1+z_n)^3	&	z_n\leq 6	\\[10pt]
		\displaystyle\sum_{i=1}^n\Delta n_e^{\rm free}(z_i)	&	z_n>6
	\end{cases}\,\,,
\end{equation}
where $n_{e0}=(1-Y_p/2)n_{b0}$ is the present-day electron density. The contribution to $\tau$ from $z\leq 6$ is $\tau_{z\leq 6}\simeq 0.038$, whereas we denote the contribution due to ALP-photon conversions at $z>6$ by $\tau_{\rm ALP}$. Therefore, by imposing that $\tau=\tau_{z\leq 6}+\tau_{\rm ALP}<\tau_{\rm exp}^{2\sigma}$, where $\tau_{\rm exp}^{2\sigma}=0.068$ is the maximum allowed value (up to $2\sigma$) by Planck 2018 measurements~\cite{Aghanim:2018eyx}, we find that the contribution to optical depth due to ALP-photon conversions is bounded by
\begin{equation}
	\tau_{\rm ALP}<\tau_{\rm exp}^{2\sigma}-\tau_{z<6}\simeq 0.030\,.
\end{equation}

\begin{figure}
	\centering
	\includegraphics[width=.8\textwidth]{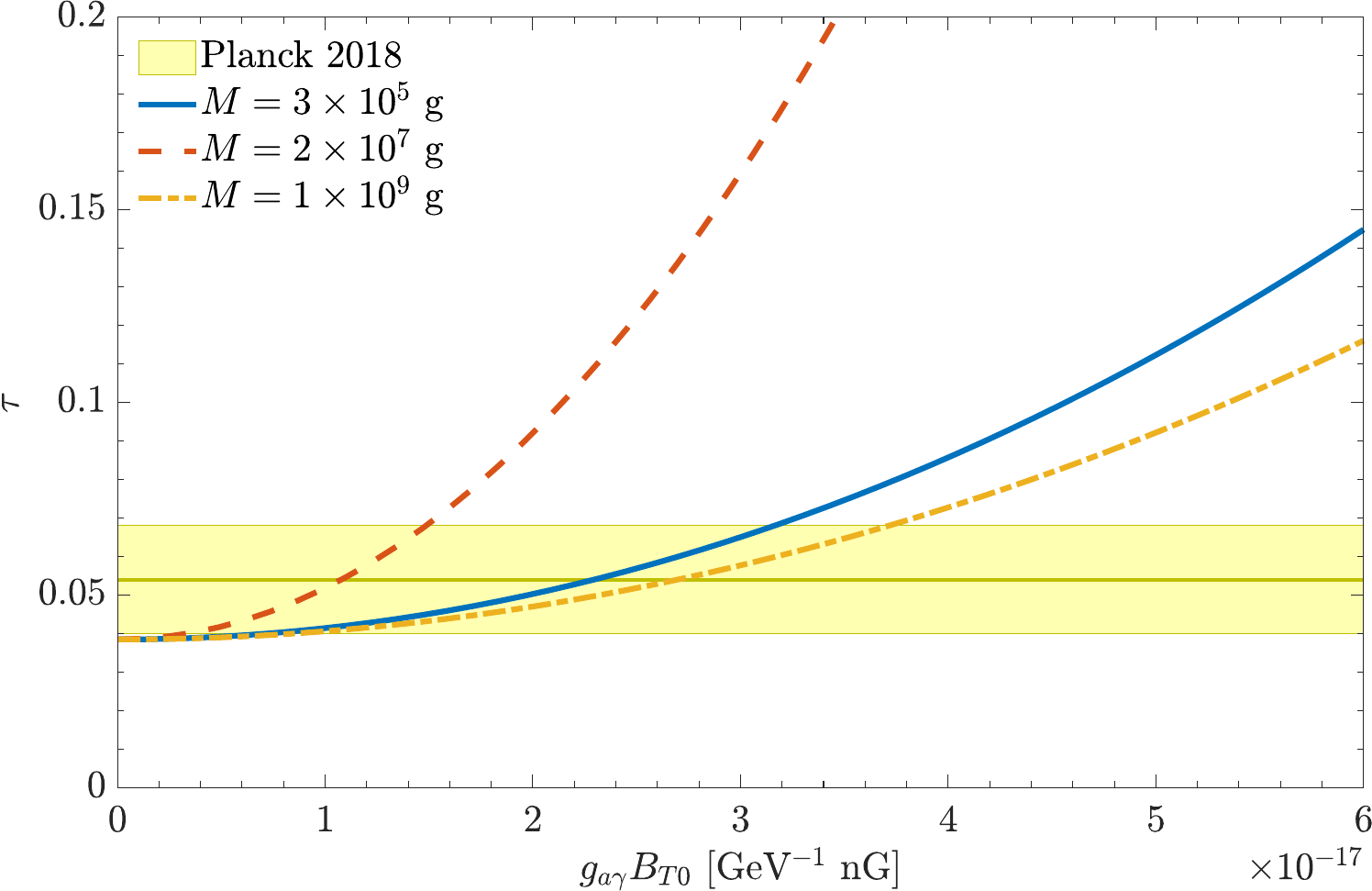}
	\caption{Plot of the Thomson optical depth $\tau$ as a function of $g_{a\gamma} B_{T0}$ for different values of the PBH mass $M$, compared with the value measured by Planck 2018.}
	\label{fig:opt_depth}
\end{figure}

In Fig.~\ref{fig:opt_depth}, we have plotted $\tau$ as a function of $g_{a\gamma} B_{T0}$ for the same PBH mass values considered in Figure \ref{fig:ionization_alp}, and we have compared the obtained values with the Planck 2018 result $\tau_{\rm exp}$. By imposing that $\tau\leq \tau_{\rm exp}^{2\sigma}$, we were able to place bounds on the ALP-photon mixing parameter  $g_{a\gamma} B_{T0}$; Fig.~\ref{fig:conversion_bounds} shows the excluded values of  $g_{a\gamma} B_{T0}$  for each PBH mass $M$ as a shaded area in the parameter space. The general result is that the upper bound on  $g_{a\gamma} B_{T0}$  is of order 
\begin{equation}
\label{eq:reion_bound}
	g_{a\gamma} B_{T0}\lesssim \mathcal{O}({10^{-15}\mbox{--}}10^{-17}) \,\textrm{GeV}^{-1}\,\textrm{nG}\,.
\end{equation}
Taking $B_{T0}\sim {1}\,\textrm{nG}$, as suggested by the present upper limits from CMB measurements~\cite{Ade:2015cva}, we find the constraint $
g_{a\gamma}\lesssim 10^{-17} \,\textrm{GeV}^{-1}$. Therefore, if a large-scale magnetic field $B_{T0}\sim {1}\,\textrm{nG}$ were present, our bound coming from reionization would be several orders of magnitude tighter than the currently favored ones from, e.g., CAST~\cite{Anastassopoulos:2017ftl} or SN 1987A~\cite{Payez:2014xsa}. Our result is compatible with that obtained in~\cite{Evoli:2016zhj}, where, however, ALP were assumed to be produced by decay of moduli in a string-theoretical scenario.

\section{Bounds on  massive ALPs decaying into photons}

In the previous Sections we have considered  astrophysical and cosmological signatures of the emission of light ALPs from an initial population of evaporating PBHs, and their conversion into photons in cosmic magnetic fields. However, if these particles are massive enough,
there exists the possibility that they spontaneously decay into photons even in the absence of an external field. It is therefore worthwhile to extend our previous results to the case of ALPs with a non-negligible mass.
The ALP spontaneous radiative decay rate is given by~\cite{Cadamuro:2011fd}
\begin{equation}
\label{eq:decay_rate}
	\Gamma_{a\gamma}=\frac{m_a^3 g_{a\gamma}^2}{64\pi}\simeq 7.55\times10^{-40}
	\left(\frac{m_a}{{1}\,\textrm{eV}}\right)^3
	\left(\frac{g_{a\gamma}}{10^{-17} \,\textrm{GeV}^{-1}}\right)^2\,\textrm{s}^{-1}\,.
\end{equation}

Different cosmological and astrophysical bounds have been discussed in Ref.~\cite{Cadamuro:2011fd} for ALPs with masses larger that $m_a \gtrsim 10$~eV, assuming a primordial thermal population of ALPs. 

The spectrum of photons produced by the ALP decay follows the Boltzmann equation 
\begin{equation}
\left[\frac{\partial}{\partial t} -H(t)\omega_\gamma\frac{\partial}{\partial \omega_\gamma}+2H(t)\right] \frac{dn_\gamma}{d\omega_\gamma}(\omega_\gamma,t)=
-\Gamma(\omega_\gamma) \frac{dn_\gamma}{d\omega_\gamma}(\omega_\gamma,t)+\frac{d^2 n_{a\gamma}(\omega_\gamma,t)}{d \omega_\gamma dt}\, ,
\label{eq:gammaevol}
\end{equation}
where $\Gamma(\omega_\gamma)$ accounts for photon absorption while the second term on the right-hand side  is the photon production rate due to ALP decay. The photon spectrum from an ALP decay with momentum $k$ and energy $\omega=[k^2+m_a^2]^{1/2}$ is flat: $dN_\gamma/d\omega_\gamma=2/k$ if $2\omega_\gamma\in I_k=[\omega-k,\omega+k]$, and zero otherwise. The contribution to photon production is thus given by
\begin{equation}
	\frac{d^2 n_{a\gamma}(\omega_\gamma,t)}{d \omega_\gamma dt}=2\int_{k_{\rm min}(\omega_\gamma)}^{+\infty}\frac{d n_a(k,t)}{dk} 
	\Gamma_{a\gamma}(\omega)\frac{dk}{k}\, ,
	\label{eq:gammaproduction}
\end{equation}
where $d n_a/dk(k,t)$ is the ALP spectrum at time $t$ and $\Gamma_{a\gamma}(\omega)=\Gamma_{a\gamma}\times m_a/\omega$, where the term $m_a/\omega$ accounts for the Lorentz boost in the decay constant.  The integration limit in Eq.~\eqref{eq:gammaproduction} can be obtained imposing that $2\omega_\gamma\in I_k$. After a simple algebra we obtain $k_{\rm min}=|\omega_\gamma-m_a^2/4\omega_\gamma|$. Finally, the evolution of the ALP spectrum is given by
\begin{equation}
\left[\frac{\partial}{\partial t} -H(t)k\frac{\partial}{\partial k}+2H(t)\right] \frac{dn_a}{dk}(k,t)=
-\Gamma_{a\gamma}(\omega)\frac{dn_a}{dk}(k,t) \, .
\label{eq:gammaevol}
\end{equation}
Although a formal solution of these equation can be obtained \cite{Masso:1999wj}, for practical purposes it is convenient to follow the approach of Sec.~5 and discretize the time/redshift. The photon spectrum at redshift $z_{n+1}=z_n+\delta z_n$ can be written as
\begin{eqnarray}
\frac{d n_\gamma(\omega_\gamma,z_{n+1})}{d \omega_\gamma}&\simeq&
\left(\frac{1+z_{n+1}}{1+z_n}\right)^2\frac{d n_\gamma}{d \omega_\gamma}\left(\omega_\gamma\frac{1+z_n}{1+z_{n+1}},z_n\right)e^{-\Gamma_n(\omega_\gamma) \delta t_n}\nonumber \\
&+&2\int_{k_{\rm min}}^{+\infty}\frac{d n_a(k,z_n)}{dk} \left(1-e^{-\frac{m_a \Gamma_{a\gamma}}{\omega}\delta t_n}\right)\frac{dk}{k}\, ,
\label{eq:photonevol}
\end{eqnarray}
with 
\begin{equation}
\frac{d n_a(k,z_{n+1})}{dk}=\left(\frac{1+z_{n+1}}{1+z_n}\right)^2\frac{d n_a}{dk}\left(k\frac{1+z_n}{1+z_{n+1}},z_n\right)e^{-\frac{m_a \Gamma_{a\gamma}}{\omega}\delta t_n}\, .
\label{eq:evolalpdens}
\end{equation}
In the previous expressions we can relate $\delta t_n$ to $\delta z_n$ through the relation 
\begin{equation}
\delta t_n = \frac{1}{H(z_n)}\frac{\delta z_n}{1+z_n}\, .
\end{equation}

\begin{figure}
\centering
\includegraphics[width=.82\textwidth]{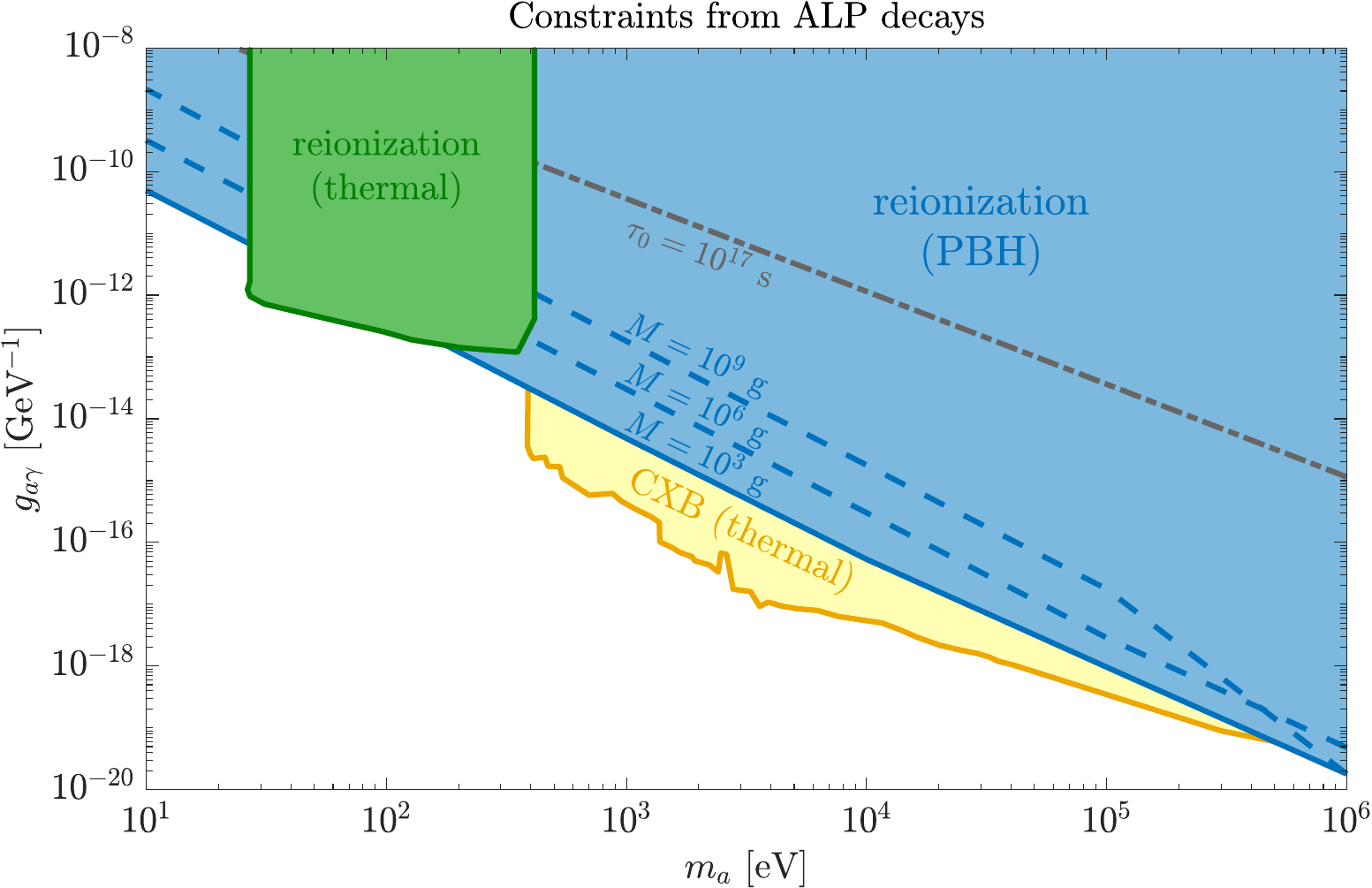}
\caption{
Exclusion plot in the $g_{a\gamma}-m_a$ plane for decaying ALPs from PBH. The bounds from reionization are given for three values of the PBH mass ($M=10^3, 10^6, 10^9$~g).
These bounds are compared to cosmological ones for a thermal population of ALPs reported in Ref.~\cite{Cadamuro:2011fd}. A line corresponding to an ALP lifetime comparable to the age  of the Universe, $\tau_0=10^{17}$~s, is also shown for comparison.
}
\label{fig:decay_bounds}
\end{figure}

The photons produced by decays would go on to reionize the intergalactic medium. The free electron fraction can be calculated by means of Eq.~\eqref{eq:D_free_elec} and the optical depth with Eq.~\eqref{eq:opt_depth}. {The effect of ALP decay on reionization, which is qualitatively different from that of ultralight ALP-photon conversions, is shown in Fig. \ref{fig:ionization_alp} for a suitable choice of parameters.}
In Fig.~\ref{fig:decay_bounds} we report these bounds as shaded areas in the $(m_a,g_{a\gamma})$ parameter space for different values of the PBH mass in the range $M=10^3-10^9$~g. Also superimposed are shown the bounds coming from reionization and non-observation of a CXB coming from decay of a thermal population of ALPs reported in Ref.~\cite{Cadamuro:2011fd}.

The results are strongly dependent on the ALP mass. Very massive ALPs ($m_a\gtrsim$~keV) are mostly non relativistic at the time of reionization thus producing photons of energy $\omega_\gamma\lesssim m_a/2$. More massive ALPs produce more energetic photons and thus a slower decay rate (in turn, a lower value of $g_{a\gamma}$) is required to produce an efficient reionization. For lighter ALP masses photons are mostly relativistic. Although the decay time is enhanced by the boost factor, ALPs can decay into photons with energies $\omega_\gamma>m_a/2$ and the two effects compensate each other. 
{For comparison we also show where in the parameter space the ALP decay lifetime $\tau_0$ would be compatible with  corresponding to the age  of the Universe $\tau_0=10^{17}$~s.} We realize that 
for the values of $g_{a\gamma}$ we are considering, ALPs would be stable on cosmological time-scales and thus would behave as dark radiation.

Although in principle decaying ALPs can give rise to a cosmic X- and $\gamma$-ray background, we have verified that the photon flux produced by ALP decay is always much smaller than the diffuse background flux in~\cite{Ballesteros:2019exr} within the bounds imposed by reionization.

\section{Conclusions}

We considered a cosmological scenario in which PBHs form with a very small initial abundance,  dominate the energy density of the Universe for a transient period before BBN, and evaporate before this epoch via Hawking radiation. We assume that the latter is composed not only 
 by Standard Model particles, but also by ALPs coupled with photons.
 For masses $m_a \lesssim 10$~MeV and photon-ALP couplings as small as the ones considered above
 these ALPs would be stable on cosmological scales and thus might appear as  dark radiation of the Universe with a contribution to the effective number
 of neutrinos {$\Delta N_{\rm eff} \simeq 0.06$}, in the reach of sensitivity of future satellite experiments like CMB-S4~\cite{Baumann:2015rya}. 
 Furthermore, for ultralight ALPs, the presence of cosmic magnetic fields would trigger sizable conversions of these ALPs into X-rays leading to possible observable signatures of this scenario.  Notably, ALP conversions might explain the soft X-ray excess observed in the Coma cluster.
 Constraints can also be placed from the cosmic X-ray background flux and from the effect on the primordial reionization. 
 We report these latter bounds in Fig.~\ref{fig:conversion_bounds} compared with those  placed by CAST experiment search of solar ALPs, by the absence of gamma-rays in coincidence with the SN 1987A and by the analysis of CMB distortions. 
 We realize that assuming a primordial magnetic field with actual value $B_{T0}=1$~{nG} one expects to improve the current constraints 
 on $g_{a\gamma}$ by different orders of magnitude. Intriguingly, such small values of $g_{a\gamma}$ are in the reach of the planned ABRACADABRA experiment for ultralight ALP 
 dark matter~\cite{Kahn:2016aff}.
 Massive ALPs (with masses $m_a \gtrsim 10$~eV) can also give an imprint on reionization via the photon flux produced by their radiative
 decays, as shown in the bounds reported in Fig.~\ref{fig:decay_bounds}.

 In principle one would expect a similar phenomenology  if gravitons  were emitted as Hawking radiation.
 However, due to the smallness of the graviton-photon coupling, given by the inverse of the Planck mass, the conversion effects would produce  marginal effects~\cite{Dolgov:2011cq,Dolgov:2013pwa}.  Nonetheless the case of rotating PBHs, where an enhancement of the graviton flux has been predicted~\cite{Hooper:2020evu}, remains an interesting case to be explored.

Future cosmological and  astrophysical observations  as well as direct ALP searches would be fundamental to probe the cosmological ALP background. 
Measurements of a positive $\Delta N_{\rm eff}$ together with a X-ray signal observed 
in Galaxy Clusters would point towards a non-thermal ALP production. In previous literature, ALPs produced by moduli decays in the early Universe have been discussed and a similar phenomenology has been discussed. However, in principle the two scenarios might be distinguished through a spectral analysis of
the produced X-rays. Furthermore the evaporation of PBHs with $M \lesssim 10^{9}$~g would lead to detectable density fluctuations induced by gravitational waves~\cite{Domenech:2021wkk}.
This intriguing scenario confirms once more the fundamental role of the elusive ALPs as cosmic messengers offering new opportunities to learn about the early Universe.

\section*{Acknowledgments}
We warmly thank Igor Irastorza, Maurizio Giannotti and Isabella Masina for reading the manuscript and for their useful comments on it.
{F.S. also acknowledges Maurizio Gasperini for the useful discussions during the development of this work.}
The work of A.M. and D.M.   is partly supported by the Italian Ministero dell’Universit\`a e Ricerca (MUR) through the research grant no. 2017W4HA7S ``NAT-NET: Neutrino and Astroparticle Theory Network'' under the program PRIN 2017, and by the Istituto Nazionale di Fisica Nucleare (INFN) through the ``Theoretical Astroparticle Physics'' (TAsP) project. The work of F.C. is supported by the U.S. Department of Energy under the award number DE-SC0020250 and DE-SC0020262.


\end{document}